\documentclass[10pt]{article}

\usepackage{hyperref}
\usepackage{authblk}
\usepackage[toc,page]{appendix}

\usepackage{graphicx}%
\usepackage{multirow}%
\usepackage{makecell} 
\usepackage{amsmath,amssymb,amsfonts}%
\usepackage{amsthm}%
\usepackage{mathrsfs}%
\usepackage{xcolor}%
\usepackage{textcomp}%
\usepackage{manyfoot}%
\usepackage{booktabs}%
\usepackage{listings}%
\usepackage{slashed}

\usepackage{pifont}
\newcommand{\xmark}{\ding{55}}%

\newcommand{\vol}{v_4}

\newcommand{\FF}{\mathbb{F}}

\newcommand{\se}{\slashed{e}}

\newcommand{\AAA}{\mathbb{A}}

\newcommand{\ussvaldivia}{Facultad de Ingenier\'ia, Universidad San Sebasti\'an, General Lagos 1163, Valdivia, Chile}
\newcommand{\ussconce}{Facultad de Ingenier\'ia, Universidad San Sebasti\'an, Lientur 1457, Concepci\'on, Chile}
\newcommand{\cecs}{Centro de Estudios Cient\'ificos, Universidad San Sebasti\'an, Avenida Arturo Prat 514, Valdivia, Chile}

\begin{document}

\title{A Geometric Action Principle for Unification of Yang-Mills Fields, Gravity and Matter Fermions}

\author[1,3]{Pedro D. Alvarez \thanks{E-mail \href{mailto:pedro.alvarez@uss.cl}{\nolinkurl{pedro.alvarez@uss.cl}}}}

\author[2]{Fernando Izaurieta \thanks{E-mail \href{mailto:fernando.izaurieta@uss.cl}{\nolinkurl{fernando.izaurieta@uss.cl}}}}

\author[2]{Cristian Quinzacara \thanks{E-mail \href{mailto:cristian.quinzacara@uss.cl}{\nolinkurl{cristian.quinzacara@uss.cl}}}}

\affil[1]{\ussvaldivia}
\affil[2]{\ussconce}
\affil[3]{\cecs}

\date{}
\maketitle

\begin{abstract}
A single geometric action, constructed as a linear combination of a restricted six-dimensional family of Clifford-generated invariant terms, simultaneously generates the gravitational, Yang--Mills, and fermionic sectors and correlates their structure and relative coefficients. The resulting parameter space contains known linearized ghost-free quadratic gravitational models with propagating torsion in a reduced bosonic sector. Our results therefore establish an economical geometric route to gauge--gravity unification that requires neither extra dimensions nor an externally introduced symmetry-breaking field. In particular, the Clifford-algebraic construction naturally accommodates parameter subspaces in which the reduced bosonic sector reproduces known ghost-free theories and the independent Klein--Gordon-type fermionic kinetic term vanishes.
\end{abstract}


\section{Introduction}\label{sec1}

In this work, we introduce a Clifford--algebraic invariant action that leads to an explicit and unified formulation of gravity, Yang--Mills interactions, and fermionic matter, all expressed in terms of a single gauge curvature. Remarkably, this geometric construction is achieved without invoking extra spacetime dimensions or an externally introduced symmetry-breaking field, while the physically admissible sectors are selected by algebraic relations among the coefficients of the selected family of terms.

The long--standing program of unifying the fundamental interactions in theoretical physics naturally comprises two complementary objectives. The first is the embedding of the gauge structure of the Standard Model (SM) into a single, coherent algebraic framework. The second is the reformulation of gravity as a genuine gauge theory, on the same conceptual footing as the remaining interactions. Each of these goals has proven highly nontrivial when pursued independently, and their simultaneous realization within a single, consistent framework has remained an open challenge. The construction presented here provides a single, coherent gauge--theoretic setting to  address both aspects of the problem.

Grand unified theories (GUTs) embed the SM group $SU(3)\times SU(2)\times U(1)$ into a simple symmetry such as $SU(5)$, $SO(10)$, or $E_6$~\cite{GeorgiGlashow1974,Georgi:1999wka}. The chiral spinor representation of $SO(10)$ elegantly accommodates one SM family plus a right-handed neutrino in a single 16-dimensional representation. Larger exceptional groups further extend this structure, linking to neutrino mass generation and charge quantization, but face persistent phenomenological issues such as proton decay and doublet--triplet splitting. In parallel, algebraic programs have explored whether the SM gauge and matter content could emerge naturally from deeper mathematical structures such as division algebras and Clifford modules. In particular, the octonions $\mathbb{O}$, \emph{whose automorphism group is $G_2$}, connect to exceptional Lie algebras through the Freudenthal--Tits magic square and encode one SM generation's quantum numbers within minimal ideals of $\mathbb{C}\!\otimes\!\mathbb{H}\!\otimes\!\mathbb{O}$~\cite{Furey2018,Furey2016,Dixon1994,GurseyTze1996,Baez2002}. These algebraic schemes are mathematically rigid and aesthetically appealing, yet a fully predictive model encompassing dynamics, symmetry breaking, and fermion masses remains unrealized.

In parallel, gauge-theoretic formulations of gravity recast Einstein's theory in first-order form using the tetrad $e^a{}_{\mu}$ and spin connection $\omega^{ab}{}_{\mu}$~\cite{Palatini:1919ffw}. Early works by Utiyama, Sciama, Kibble, Hayashi and Nakano, and later Hehl, established that while local Lorentz symmetry can be treated as a gauge group, the resulting actions differ from Yang--Mills theories because the tetrad does not transform as a connection~\cite{Utiyama:1956sy,Sciama:1964wt,Kibble:1961ba,Hayashi:1967se,Hehl:2023khc,Krasnov:2020lku}. Attempts to extend the gauge principle to the full Poincar\'e group face degeneracies in the algebra's Cartan--Killing metric that render translational components non-dynamical~\cite{Kawai:1979ta,Aldrovandi:1981mv}. Weyl's early generalization of the Dirac equation to curved spacetime showed that spinors intrinsically require a Lorentz gauge structure~\cite{Weyl:1929fm}. Later, the MacDowell--Mansouri construction framed gravity as a symmetry-breaking gauge theory with an algebraic dual operator selecting the Lorentz-invariant sector~\cite{MacDowell:1977jt,Wise:2006sm}.

The repeated emergence of Clifford algebras across grand unification, algebraic models, and gauge formulations of gravity suggests a common geometric foundation. Building on this observation, previous work~\cite{Alvarez:2013tga,Alvarez:2021zsw,Alvarez:2022eew} proposed that implementing supersymmetry in the adjoint representation yields unified actions of the form
\begin{equation}
 S = \int \langle\, \FF \circledast \FF \,\rangle ,\label{USUSY-MMaction}
\end{equation}
where $\FF$ is the field strength and $\circledast$ a generalized dual operator~. However, that construction defined the action of $\circledast$ only on two-forms in four dimensions, limiting its scope. This is a significant drawback of the $\circledast$-setup that we circumvent by developing the usage of Clifford-algebra--valued frames presented in this paper. In other words, anti-symmetric products of Clifford-algebra--valued frames naturally reproduce existing invariant structures via the Hodge dual but also can reproduce new invariant structures that are included in the superalgebra but excluded by the usage of $\circledast$-only, as in \eqref{USUSY-MMaction}.

A major obstacle to unification is that naive Yang--Mills-type formulations of gravity generically produce higher-derivative curvature invariants and torsion terms that lead to ghost instabilities. This obstruction is not accidental but structural. We present an approach to circumvent this problem allowing us to construct a single master Lagrangian built from a Clifford-algebra--valued curvature and a selected family of six parity-even four-form gauge invariants that simultaneously reproduce the canonical kinetic terms of gravity, Yang--Mills fields, and spin-$\tfrac{1}{2}$ fermions.

Remarkably, within the restricted space of quadratic invariants generated by this construction, the reduced bosonic sector contains known ghost-free quadratic gravity theories with propagating torsion, originally identified at the linearized level by Sezgin. Within the same restricted parameter space, the independent Klein--Gordon--type fermionic kinetic term vanishes on an algebraically defined codimension-one locus, while the remaining antisymmetric two-derivative structure can be rewritten as nonminimal curvature and torsion couplings.

This is analogous in spirit to other geometric gauge constructions, such as Chern--Simons-type invariants and MacDowell--Mansouri-type actions, where specific relative coefficients are correlated by the underlying geometric structure rather than by phenomenological tuning of different sectors independently. Before proceeding to the details of the construction, let us also clarify that the present model is not simply a rewriting of Chern--Simons or MacDowell--Mansouri-type theories, since it contains terms that are not present in those frameworks.

\section{Yang--Mills Gravity in 4D: The Obstruction}

The first obstacle to unifying gravity with the Yang-Mills interactions are the structural differences among them. On the one hand the Yang--Mills Lagrangian constitutes the dynamical core of the Standard Model,
\begin{equation}
	L_{\mathrm{YM}}
	=
	-\frac{1}{2}\,\big\langle \mathbb{F}\wedge \ast \mathbb{F}\big\rangle ,
\end{equation}
where the gauge curvature two--form is defined as
\footnote{For algebra--valued $p$- and $q$-forms, the graded commutator is given by
	$\left[\mathbb{P},\mathbb{Q}\right]
	=\mathbb{P}\wedge\mathbb{Q}-(-1)^{pq}\mathbb{Q}\wedge\mathbb{P}
	=P^{A}\wedge Q^{B}C_{AB}{}^{C}\boldsymbol{T}_{C}$.}
\begin{equation}
	\mathbb{F}=\mathrm{d}\mathbb{A} + \mathbb{A} \wedge \mathbb{A}\,.
\end{equation}
Here $\langle\,\cdot\,,\,\cdot\,\rangle$ denotes an invariant quadratic form on the Lie algebra, and $$\ast:\Omega^{p}\to\Omega^{d-p}$$ is the Hodge dual. The latter implicitly introduces a metric structure on spacetime, or equivalently a vierbein one--form $e^{a}=e^{a}{}_{\mu}\mathrm{d}x^{\mu}$,
which relates the spacetime metric to the Minkowski metric through $g_{\mu\nu}=\eta_{ab}e^{a}{}_{\mu}e^{b}{}_{\nu}$.

On the other hand, the Einstein--Hilbert Lagrangian for gravity
\begin{equation}
	L_{\mathrm{EH}}=
	\frac{1}{4}\,\epsilon_{abcd}\,R^{ab}\wedge e^{c}\wedge e^{d},
\end{equation}
is built from superficially similar ingredients: the Lorentz gauge curvature
two--form components $R^{ab}=\mathrm{d}\omega^{ab}+\omega^{a}{}_{c}\wedge\omega^{cb}$, the invariant tensor $\epsilon_{abcd}$, and the vierbein one--form. Nevertheless, the two constructions are fundamentally different. In particular, a naive attempt to formulate gravity as a Yang--Mills theory fails, though in a manner that is highly instructive.

A close parallel between both theories is obtained by considering the spin-$\tfrac{1}{2}$ representation of the AdS algebra realized through gamma matrices,
\begin{equation}
	\boldsymbol{\Sigma}_{ab}
	=
	\frac{1}{2}\gamma_{ab}
	=
	\frac{1}{4}\left[\gamma_{a},\gamma_{b}\right],
	\qquad
	\boldsymbol{P}_{a}\label{Eq_Rep_s1/2_AdS}
	=
	\frac{1}{2}\gamma_{a},
\end{equation}
and introducing an AdS--valued gauge connection,
\begin{equation}
	\mathbb{A}=
	\frac{1}{2}\omega^{ab}\boldsymbol{\Sigma}_{ab}
	+e^{a}\boldsymbol{P}_{a}.
\end{equation}
The associated curvature two--form is then
\begin{equation}
	\mathbb{F}
	=
	\frac{1}{2}\left(R^{ab}+e^{a}\wedge e^{b}\right)\boldsymbol{\Sigma}_{ab}
	+T^{a}\boldsymbol{P}_{a},\label{Eq_F_AdS}
\end{equation}
where $T^{a}=\mathrm{d}e^a + \omega^a{}_b\wedge e^b$ denotes the torsion two--form.

The corresponding AdS Yang--Mills Lagrangian reads
\begin{equation}
	\begin{aligned}
		-\frac{1}{2}\left. \big\langle \mathbb{F}\wedge \ast \mathbb{F}\big\rangle\right | _{\mathrm{AdS}}
		=&
		\frac{1}{4}\epsilon_{abcd}\,R^{ab}\wedge e^{c}\wedge e^{d}
		+\frac{1}{8}\epsilon_{abcd}\,e^{a}\wedge e^{b}\wedge e^{c}\wedge e^{d}
		\\[2pt]
		&+\frac{1}{4}R^{ab}\wedge\ast R_{ab}
		-\frac{1}{2}T^{a}\wedge\ast T_{a}.
	\end{aligned}
\end{equation}

The first two terms reproduce the Einstein--Hilbert term and a cosmological constant contribution. The remaining terms correspond to the Kretschmann invariant and a quadratic torsion contraction, rendering the theory generically ghost-ridden. This general problem is precisely the obstruction addressed in the MacDowell--Mansouri construction, where the Hodge dual is replaced by a new generalized dual mapping the Kretschmann into the Gauss--Bonnet density. While effective, this prescription is intrinsically \emph{ad hoc}, as no principled criterion singles out the generalized dual over the standard Hodge dual on the Lorentz algebra.

Circumventing this obstruction naturally motivates the search for admissible
invariants beyond the standard Yang--Mills construction.

\section{Clifford-Valued Gauge Fields and Admissible Invariants}

The problem analysed in the last section naturally raises the question of whether, using the same basic ingredients as in Yang--Mills theory, it is possible to construct Lagrangians beyond the classical Yang--Mills term for a generic Lie algebra. The answer is affirmative. In odd dimensions, Chern--Simons theories provide a well--known class of gauge models relying solely on the gauge structure and Lie algebra invariants, without reference to a Hodge dual (see~\cite{Zanelli:2005sa} and references therein). In four dimensions, additional possibilities beyond the YM term also arise once the role of the vierbein is properly understood. To find them, the key observation is that the inclusion of the Hodge dual already presupposes the presence of the vierbein one--form. Within a gauge--theoretic framework, the vierbein may therefore appear only through the Hodge dual or as an algebra--valued object, namely the Clifford--valued one--form
\begin{equation}
	\se \;=\; e^{a}\,\gamma_{a} \label{se},
\end{equation}
where it is assumed that the corresponding Lie algebra or Lie superalgebra admits a Clifford representation. In fact, a weaker condition is sufficient for the present construction: it is enough that the spacetime generators of the Lie superalgebra are realized in a Clifford representation.

We select a six-dimensional family of parity--even contributions of the form $\langle \mathbb{X}\wedge\ast\mathbb{X}\rangle$ that are at most quadratic in the gauge curvature. Here $\mathbb{X}$ is a Lie--algebra--valued $p$--form constructed from ordered products of $\se$ and at most one factor of $\mathbb F$.

For the curvature-dependent sector, we retain the Yang--Mills
term and the frame-inserted and commutator structures listed below.

Pure-frame expressions of the form $\langle\se^{\,p}\wedge *\se^{\,p}\rangle$ reduce, up to numerical factors and signs, to the volume form and therefore we may choose $\langle \slashed{e}\wedge\ast\slashed{e}\rangle$ as a representative of such terms.

For the curvature-dependent two--form sector, we select \[ \mathbb{X}^{(2)}=\mathbb{F}, \] which gives rise to the standard Yang--Mills invariant $\langle\mathbb{F}\wedge\ast\mathbb{F}\rangle$.

At the level of three--forms, we retain two structures,
\[
    \mathbb{X}^{(3)}_1=\slashed{e}\wedge\mathbb{F},
    \qquad
    \mathbb{X}^{(3)}_2=[\slashed{e},\mathbb{F}].
\]
Similarly, at the level of four--forms, we retain:
\[
\mathbb{X}^{(4)}_1=\slashed{e}^{\, 2}\wedge\mathbb{F},
\qquad
\mathbb{X}^{(4)}_2=[\slashed{e}^{\, 2},\mathbb{F}] .
\]
The selected family of parity--even Clifford-generated invariants is
\begin{align}
&\langle \slashed{e}\wedge  \ast\slashed{e} \rangle,\label{<sese>UT} \\
&\langle \mathbb{F}\wedge  \ast\mathbb{F} \rangle,\label{<FF>UT}\\
&\langle\slashed{e}\wedge\mathbb{F}\wedge  \ast(\mathbb{F}\wedge\slashed{e}) \rangle,\label{<seFFse>UT}\\
&\langle\slashed{e}^{\, 2}\wedge\mathbb{F}\wedge  \ast(\mathbb{F}\wedge\slashed{e}^{\, 2}) \rangle,\label{<seseFFsese>UT}\\
&\langle\left[\slashed{e},\mathbb{F}\right]\wedge  \ast\left[\mathbb{F},\slashed{e}\right] \rangle,\label{<seFFse_v2>UT} \\
&\langle [\slashed{e}^{\, 2},\mathbb{F} ]\wedge \ast [\mathbb{F},\slashed{e}^{\, 2} ] \rangle ,\label{<seseFFsese_v2>UT}
\end{align}

The explicit evaluation of these invariants is most conveniently carried out in a superalgebra matrix representation whose spacetime block is simultaneously a representation of the Clifford algebra. In the $osp(n|4)$ realization used below, the Clifford generators are embedded as $\Gamma_a=\operatorname{diag}(\gamma_a,0)$, so that the frame is itself lifted to the superalgebra according to $\se=e^a\Gamma_a=2e^a\boldsymbol{P}_a$ and $\se^2=e^a\wedge e^b\,\Gamma_a\Gamma_b$. The brackets in Eqs.~\eqref{<sese>UT}--\eqref{<seseFFsese_v2>UT} can then be evaluated directly in terms of invariant supertraces. The Yang--Mills and commutator structures are controlled by quadratic invariant tensors, whereas the frame-inserted terms probe ordered quartic invariant tensors such as $\langle\Gamma_a\boldsymbol{T}_A\boldsymbol{T}_B\Gamma_b\rangle$ and $\langle\Gamma_{ab}\boldsymbol{T}_A\boldsymbol{T}_B \Gamma_{cd}\rangle$. Consequently, the reduction to ordinary tensor components involves only supermatrix multiplication, Clifford-algebra identities, and standard Hodge contractions. The explicit calculation, including the gravitational, Yang--Mills, and fermionic sectors, is given in Appendix~\ref{app:table1-derivation}.

The four-forms~\eqref{<sese>UT}--\eqref{<seseFFsese_v2>UT} make explicit that the standard Yang--Mills term is not isolated, but belongs to a broader finite family of Clifford-generated structures. In four spacetime dimensions, contributions containing more than four explicit factors of $\slashed e$ vanish by form-degree nilpotency.

Of course, one may construct parity--odd invariants, extend the analysis to higher dimensions, or consider higher--order curvature terms. These generalizations will be discussed elsewhere. The central result of the present work, however, is that the six parity--even invariants \eqref{<sese>UT}--\eqref{<seseFFsese_v2>UT} suffice to define the six-parameter master Lagrangian studied in this work.

Explicitly, the general linear combination within the selected six-dimensional family is
\begin{align}
	L_{\mathrm{LL}}\left(\mathbb{A}\right)
	&=  \alpha \langle \mathbb{F}\wedge \ast\mathbb{F} \rangle +\beta \langle \slashed{e}\wedge\mathbb{F}\wedge \ast(\mathbb{F}\wedge\slashed{e})  \rangle +\gamma \langle \slashed{e}^{\, 2}\wedge\mathbb{F}\wedge \ast(\mathbb{F}\wedge\slashed{e}^{\, 2})  \rangle \nonumber\\
	&+\delta \langle \left[\slashed{e},\mathbb{F}\right]\wedge \ast [\mathbb{F},\slashed{e} ] \rangle +\varepsilon \langle [\slashed{e}^{\, 2},\mathbb{F} ]\wedge \ast [\mathbb{F},\slashed{e}^{\, 2} ]\rangle +\zeta \langle \slashed{e}\wedge \ast\slashed{e} \rangle\,, \label{LLlagrangian}
\end{align}
where we also included a term of zero order in the curvature.

The main result of the present paper is the following: in four spacetime dimensions, the parity--even Clifford--generated invariants \eqref{<sese>UT}--\eqref{<seseFFsese_v2>UT} define a restricted finite-dimensional family of Yang--Mills--type Lagrangians that are at most quadratic in the gauge curvature. Their general linear combination, Eq.~\eqref{LLlagrangian}, provides the master four--form action studied here. Once the Lorentz algebra is embedded into a suitable superalgebra, this construction simultaneously generates gravity, Yang--Mills interactions, and spin-1/2 fermions within a common gauge--theoretic framework in $d=4$, without the introduction of extra dimensions, auxiliary fields, or \emph{ad hoc} symmetry breaking of the Stelle-West type \cite{Stelle:1979aj}. The details of the tensorial expression of \eqref{LLlagrangian} can be found in the appendix \ref{app:table1-derivation}, see formula \eqref{eq:app-master-paper}.

It is important to emphasize that the Clifford-valued frame $\se=e^{a}\gamma_{a}$ transforms covariantly under the bosonic gauge symmetries that remain manifest in the present formulation, namely local Lorentz transformations and the internal $SO(n)$ gauge symmetry. Consequently, each of the invariants \eqref{<sese>UT}--\eqref{<seseFFsese_v2>UT}, and hence any action constructed as their linear combination, is invariant under these bosonic gauge transformations. Although the full fermionic part of the supergauge symmetry requires a separate discussion, the common embedding of the spacetime connection, the internal gauge fields, and the fermionic matter sector into a single superconnection together with Clifford-valued frame insertions is already sufficient to correlate their gauge structures through the same supercurvature. In this sense, the construction provides a unified gauge-theoretic organization of the gravitational and internal gauge bosons together with the matter fields contained in the gauge connection. The additional role of the matter ansatz will be discussed below, see \eqref{matteransatz}. Its implications for the residual supersymmetries have been discussed in \cite{Alvarez:2021qbu,Alvarez:2021zhh}.

A technical remark is in order. The Clifford--exterior algebra generated by the frame is closed under the action of the Hodge dual,
\begin{equation}
    \gamma_{\ast}\,\ast\slashed{e}^{\, p}
    = \frac{p!}{(d-p)!}(-1)^{\frac{1}{2}p(p+1)}\,\slashed{e}^{\,d-p},\label{hodgesese}
\end{equation}
with $\gamma_{\ast}=\gamma_{0}\gamma_{1}\gamma_{2}\gamma_{3}$. In the present analysis we nevertheless retain the standard Hodge dual in terms~\eqref{<sese>UT}--\eqref{<seseFFsese_v2>UT}, so as to avoid obscuring the conceptual structure with additional technical machinery. We note, however, that the Hodge dual itself admits a natural formulation entirely within the Clifford--exterior algebra~\cite{Lawson:1998yr}.

In the gravitational sector, the term~\eqref{<sese>UT} corresponds purely to a cosmological constant contribution and, when the AdS curvature~\eqref{Eq_F_AdS} is employed, the terms \eqref{<FF>UT}--\eqref{<seseFFsese>UT} also generate the Einstein--Hilbert Lagrangian, quadratic contractions of the Riemann and Ricci tensors, the Gauss--Bonnet density, and quadratic torsional terms. The remaining contributions \eqref{<seFFse_v2>UT} and \eqref{<seseFFsese_v2>UT} generate curvature combinations that vanish in the torsionless limit by the algebraic Bianchi identities and therefore encode torsion-dependent information in the general Riemann--Cartan geometry.

While arbitrary choices of the coupling constants may lead to pathological degrees of freedom, the parameter space contains reduced bosonic sectors that reproduce known linearly ghost--free theories of gravity with torsion. As a particularly transparent example, the choice
\begin{equation}
\zeta=0, \quad  \beta=-\alpha,\quad \gamma=-\alpha/4,\quad \delta = 0, \quad \varepsilon = 0 \label{eq:torsional-ads-gravity}
\end{equation}
yields
\begin{equation}
	\left. L_{\mathrm{LL}}\right|_{\mathrm{AdS}}
	=
	-\frac{\alpha}{4}\, \epsilon_{abcd} \left(R^{ab}+e^{a}\wedge e^{b}\right) \wedge \left(R^{cd}+e^{c}\wedge e^{d}\right),\label{AdSaction}
\end{equation}
thereby canceling the Hodge dual in the torsionless sector and reducing the theory to standard AdS gravity in the limit of vanishing torsion, where the correct global sign corresponds to the choice $\alpha < 0$. However, it is not the only available choice for a sensible gravity theory in $d=4$. The new Lagrangian~\eqref{LLlagrangian} allows us to reproduce, within a gauge--theoretic framework, the results of the seminal analyses of Refs.~\cite{Sezgin:1979zf,Sezgin:1981xs}, where classes of parity--even quadratic Poincar\'e--gauge Lagrangians that are ghost- and tachyon-free at the linearized level were identified. A key advantage of the present approach is that it contains not only the gravitational sector, but also gauge bosons and spinors, therefore the Clifford--algebra formulation provides a natural geometric principle that contains admissible ghost--free sectors, as discussed further in Section~\ref{sec:ghost-free}.

\section{A Unified Geometric Principle for Gravity, Gauge Bosons and Fermions}

In order to incorporate internal gauge bosons and spinorial matter while evading the assumptions of the Coleman--Mandula theorem, the Lorentz algebra can be embedded into a superalgebra such as $osp(n|4)$ or $su(2,2|n)$. This enlargement necessarily introduces fermionic generators, which are naturally interpreted as supercharges \cite{Alvarez:2022eew}.

For definiteness, and to keep the exposition transparent, we restrict attention to the case
\[
osp(n|4)_{\bar 0} \simeq so(n)\oplus sp(4,\mathbb{R}) \simeq so(n)\oplus so(3,2),
\]
where ``$\bar 0$'' stands for the ``bosonic part of''. The spacetime factor admits a Clifford representation and the fermionic generators can be taken to be Majorana supercharges. This corresponds to an internal symmetry of $so(n)$ type. The same strategy suggests an analogous extension to the $su(2,2|n)$ case:
\[
su(2,2|n)_{\bar 0} \simeq su(2,2)\oplus su(n)\oplus u(1) \simeq so(4,2)\oplus su(n)\oplus u(1),
\]
but reality, $U(1)$ representation and fermionic--sector issues require a separate treatment. One key difference in this case is that the supercharges are Dirac spinors. For this reason we will focus on the simplest model that illustrates the construction defined in this paper is the $osp(n|4)$ model. The subtleties of a model of the $su(2,2|n)$-type will be discussed elsewhere.

We adopt a block supermatrix representation of the form
\begin{equation}
	\left[
	\begin{array}{c|c}
		AdS_{4\times 4} & Q_{4\times n} \\
		\addlinespace[0.2em]
		\hline
		\addlinespace[0.2em]
		Q^{T}_{n\times 4} & so(n)_{n\times n}
	\end{array}
	\right],
	\label{superalgebrarep}
\end{equation}
which makes explicit the decomposition into anti--de Sitter, fermionic, and internal sectors; see Appendix~\ref{app:table1-derivation} for more details.

Following the strategy developed in Refs.~\cite{Alvarez:2013tga,Alvarez:2021zsw,Alvarez:2022eew},  and in order to construct a genuinely geometric model, all dynamical variables are encoded into a single gauge connection valued in the superalgebra
$osp(n|4)$,
\begin{equation}
	\AAA
	\;=\;
	\frac{1}{2}\,\omega^{ab}\,\boldsymbol{\Sigma}_{ab}
	+ e^a\,\boldsymbol{P}_a
	+ \frac{1}{2}\,A^{ij}\,\boldsymbol{L}_{ij}
	+ \overline{\boldsymbol{Q}}^{\,i}_{\alpha}\,(\se \psi)^\alpha_i \, ,
	\label{gaugeconnection}
\end{equation}
thereby placing gravity, internal gauge fields, and fermionic matter within a common superconnection. Throughout this work the AdS radius is set to unity. It may be restored by replacing $e^a\boldsymbol P_a$ with $\ell^{-1}e^a\boldsymbol P_a$ in the connection and inserting the corresponding powers of $\ell^{-1}$ in the curvature and composite fermionic terms.
The generators $\boldsymbol{\Sigma}_{ab}$ and $\boldsymbol{P}_a$ are those of the AdS algebra in the spin-$\tfrac{1}{2}$ representation introduced in Eq.~\eqref{Eq_Rep_s1/2_AdS}, with the understanding that in the superalgebra context
\begin{equation}
	\Gamma_a =
	\left[
	\begin{array}{c|c}
		\gamma_a & 0_{4\times n} \\
		\addlinespace[0.2em]
		\hline
		\addlinespace[0.2em]
		0_{n\times 4} & 0_{n\times n}
	\end{array}
	\right];
\end{equation}
so, when embedded in the super algebra, the Clifford-valued frame is to be understood as $\se=e^{a}\Gamma_a$. The generators $\overline{\boldsymbol{Q}}^{\,i}_{\alpha}$ correspond to Majorana supercharges, while $\boldsymbol{L}_{ij}$ generate the internal $so(n)$ symmetry acting on them.

The fermionic matter sector is encoded in the composite one--form $\se\psi$, with spinor indices suppressed whenever no ambiguity arises. Explicitly,
\begin{equation}
	(\se \psi)^\alpha_i \;=\; e^a{}_{\mu}\mathrm{d}x^{\mu}\, (\gamma_a)^\alpha{}_\beta\,\psi^\beta_i .\label{matteransatz}
\end{equation}
Since $\se\psi$ is annihilated by the spin-$\tfrac{3}{2}$ projector, $\se\psi$ lies purely spin-$\tfrac{1}{2}$ degrees of freedom. The matter ansatz $\chi_i = \se\psi_i$ is imposed at the level of the action, before performing the variation \cite{Alvarez:2013tga}.

Consequently, the reduced Euler--Lagrange equations are obtained by pulling back the unrestricted variational one--form to the ansatz surface. They are therefore the projections of the unrestricted field equations along variations tangent to that surface and need not reconstruct the complete unrestricted equations. The ansatz is equivariant under the residual local $SO(1,3)\times SO(n)$ symmetry and compatible with spacetime diffeomorphisms, but it is not preserved by arbitrary local $OSp(n|4)$ transformations. This does not preclude particular solutions from possessing an enhanced symmetry: suitable bosonic backgrounds may retain the full AdS isometry group and may also preserve a subset of rigid supersymmetries whenever they admit appropriate Killing or twistor spinors \cite{deMedeiros:2012sb,Festuccia:2011ws}.

The corresponding field strength decomposes according to the same algebraic
splitting,
\begin{equation}
	\FF
	\;=\;
	\frac{1}{2} F^{ab}\boldsymbol{\Sigma}_{ab}
	+ F^a \boldsymbol{P}_a
	+ \frac{1}{2} F^{ij}\boldsymbol{L}_{ij}
	+ \overline{\boldsymbol{Q}}^{\,i}_{\alpha} X^\alpha_i ,
	\label{fieldstrength}
\end{equation}
with $X^\alpha_i$ the fermionic curvature two-form component.

Substituting~\eqref{fieldstrength} into the unified Lagrangian $L_{\mathrm{LL}}$~\eqref{LLlagrangian}, a remarkable simplification occurs in the internal $so(n)$ sector. Owing to the representation \eqref{superalgebrarep}, the structures \eqref{<seFFse>UT}--\eqref{<seseFFsese_v2>UT} vanish identically when evaluated on the internal curvature $\tfrac{1}{2}F^{ij}\boldsymbol{L}_{ij}$.
Consequently, the only surviving internal contribution arises from the term $\langle\FF\wedge\ast\FF\rangle$. Since the invariant bilinear form $\langle\cdot,\cdot\rangle$ corresponds to the supertrace in the chosen representation, one finds
\begin{align}
	\left. L_{\mathrm{LL}}\right|_{so(n)}
	&=\left.\alpha\langle\FF\wedge\ast\FF\rangle\right|_{so(n)}\\
	&=-\alpha\,\left.\mathrm{Tr}\!\left(\FF\wedge\ast\FF\right)\right|_{so(n)} .
\end{align}
With the generator normalization adopted here and setting the internal gauge coupling to unity, canonical normalization of the Yang--Mills fields corresponds to
\begin{equation}
    \alpha=-1.\label{Eq_alpha_1/2_by_YM}
\end{equation}

We now turn to the fermionic kinetic sector. The fermionic curvature $\overline{\boldsymbol{Q}}^{\,i}_{\alpha}X^\alpha_i$ is given by the $AdS\otimes so(n)$--covariant derivative of the composite field $\xi=\se\psi$,
\begin{equation}
X = \mathrm{d}\xi + \frac{1}{2}\,\omega^{ab}\Sigma_{ab}\wedge\xi + \frac{1}{2}\,A^{ij}L_{ij}\wedge\xi
 + \frac{1}{2}\,\se\wedge\xi . \label{adsspinorcovderivative}
\end{equation}
In \eqref{adsspinorcovderivative}, it is understood that the generators are in the spin-$\tfrac{1}{2}$ representation. Recognizing that in \eqref{adsspinorcovderivative} the first three terms define the
$SO(1,3)\otimes SO(n)$--covariant derivative
$\mathrm{D}=\mathrm{d}
+\tfrac{1}{2}\omega^{ab}\Sigma_{ab}
+\tfrac{1}{2}A^{mn}L_{mn}$, and writing $\xi=\se\psi$, one obtains
\begin{equation}
X = -\se\wedge\mathrm{D}\psi +\frac{1}{2}\se^{2}\psi +T^{a}\gamma_{a}\psi . \label{Eq_X}
\end{equation}

Evaluating the Lagrangian pieces
\eqref{<FF>UT}--\eqref{<seseFFsese_v2>UT} on the purely fermionic curvature
\begin{equation}
	\left.\FF\right|_{\mathrm{F}}
	=\overline{\boldsymbol{Q}}\,X ,
\end{equation}
one finds, with different weights, the standard\footnote{Here $\gamma^\mu=e_a{}^\mu \gamma^a$.} kinetic term $\bar{\psi}\gamma^{\mu}\mathrm{D}_{\mu}\psi - \mathrm{D}_{\mu}\bar{\psi}\gamma^{\mu}\psi$ and mass term $\bar{\psi}\psi$, as well as nonminimal couplings to torsion. In addition, two terms quadratic in derivatives arise,
\begin{align}
\mathrm{D}_{\mu}\bar{\psi}\, \gamma^{\mu\nu} \, \mathrm{D}_{\nu}\psi , \label{Eq_Term_DpsiGamma_abDpsi}\\
\mathrm{D}^{\mu}\bar{\psi}\,\mathrm{D}_{\mu}\psi . \label{Eq_Term_KleinGordon}
\end{align}

The term~\eqref{Eq_Term_DpsiGamma_abDpsi} can be shown, by integration by parts, to correspond to a nonminimal coupling between fermions and the Lorentz curvature, up to boundary contributions, see Appendix \ref{app:Dgamma2D}. By contrast, Eq.~\eqref{Eq_Term_KleinGordon} is a Klein--Gordon--type kinetic term, which is a well--known source of ghost degrees of freedom. To eliminate this additional second-order fermionic kinetic structure, we may impose an additional algebraic constraint on the Lagrangian coefficients,
\begin{equation}
	\alpha + 2\beta - 6 \gamma + 4 \delta - 12 \varepsilon = 0. \label{Eq_Coef_KleinGordon}
\end{equation}
see detailed derivation of Section~\ref{sec:fermion-form}, and also see Table~\ref{tab:SvN-ModGB}.

Since the anticommutator of the supercharges takes the form
\begin{equation}
\{ \boldsymbol{Q}_{i}^{\rho},\boldsymbol{\bar{Q}}_{\sigma}^{j}\} =
	\left(
	(\gamma^{a})^{\rho}{}_{\sigma}\,\boldsymbol{P}_{a}
	-\frac{1}{2}(\gamma^{ab})^{\rho}{}_{\sigma}\,\boldsymbol{\Sigma}_{ab}
	\right)\delta_{i}^{j}
	+\delta_{\sigma}^{\rho}\,\boldsymbol{L}^{j}{}_{i},
\end{equation}
it follows that the bosonic components of the gauge curvature two--form acquire fermion--dependent contributions. Explicitly, one finds
\begin{align}
F^{ab} &= R^{ab} + e^{a}\wedge e^{b} - e^{a}\wedge e^{b} \, \bar{\psi}\psi +\ast\!\left(e^{a}\wedge e^{b}\right)\bar{\psi}\gamma_{\ast}\psi,\label{Eq_Lorentz_Curvature}\\
F^a &= T^a - \ast\!\left(e^{a}\wedge e^{b}\right) \, \bar{\psi}\gamma_b \gamma_{\ast}\psi,\\
F^{i}{}_{j} &= f^{i}{}_{j} - e^{a}\wedge e^{b}\,\bar{\psi}^{i}\gamma_{ab}\psi_{j},
\end{align}
where the internal curvature is defined as
\begin{equation}
f^{i}{}_{j} = \mathrm{d}A^{i}{}_{j} + A^{i}{}_{k}\wedge A^{k}{}_{j}.
\end{equation}

Substituting these curvature components into the unified Lagrangian $L_{\mathrm{LL}}$, one generically obtains, up to numerical coefficients, an effective action of the schematic form
\begin{align}
	L_{\mathrm{LL}}\left(\mathbb{A}\right)= -\alpha\,\mathrm{Tr}\!\left(f\wedge\ast f\right)
	+L_\mathrm{grav}
	+L_{CC}
	+L_\mathrm{fer}
	+ L_\mathrm{NMC}\,.
\end{align}
Here $L_{\mathrm{grav}}$ denotes the gravitational sector, which reproduces known linearized ghost--free models for the particular parameter choices discussed below; $L_{CC}$ denotes the constant volume term, $L_{\mathrm{fer}}$ contains the kinetic and mass-like terms for the Majorana spinors, and $L_{\mathrm{NMC}}$ collects the nonminimal couplings between the fermions, Lorentz curvature, torsion, and internal $so(n)$ field strength.

The key point is that all these contributions originate from a common superalgebraic connection before restriction. After imposing the matter ansatz, its fermionic component is represented by the composite one--form $\se\psi$, and the manifest local gauge symmetry is reduced to $SO(1,3)\times SO(n)$. Within this residual gauge structure, gravity, Yang--Mills fields, and fermionic matter continue to be encoded through a single curvature and a common Clifford--algebraic construction.

\section{Gravity and fermionic sector}\label{sec:gravity-fermion-sector}

The purely gravitational sector is obtained by setting the fermions and the
internal gauge field to zero.  The corresponding AdS curvature is
\begin{equation}
\left.\FF\right|_{\mathrm{grav}}
=\frac12\hat R^{ab}\boldsymbol\Sigma_{ab}
+T^a\boldsymbol P_a,
\qquad
\hat R^{ab}:=R^{ab}+e^a\wedge e^b.
\label{eq:Fgrav-main}
\end{equation}
The ordered Clifford traces can be evaluated completely while keeping the
answer in differential-form notation.  For this purpose, let $\iota_a$ denote
contraction with the frame vector dual to $e^a$ and introduce the Ricci
one-form of the Lorentz component of the AdS curvature,
\begin{equation}
\mathfrak R^a:=\iota_b\hat R^{ba},
\label{eq:AdS-Ricci-one-form-main}
\end{equation}
where
\begin{align}
\mathfrak R^a
&=\frac12\hat R^{ba}{}_{cd}\,
\iota_b\!\left(e^c\wedge e^d\right)
\nonumber\\
&=\frac12\hat R^{ba}{}_{cd}
\left(\delta_b^c e^d-\delta_b^d e^c\right)
\nonumber\\
&=\hat R^{ba}{}_{bc}\,e^c
=\hat R^{ab}{}_{cb}\,e^c
=\hat R^a{}_c\,e^c,
\qquad
\hat R^a{}_c:=\hat R^{ab}{}_{cb}.
\label{eq:AdS-Ricci-one-form-derivation-main}
\end{align}
Separating the pure Lorentz part we have,
\begin{equation}
\hat R^a{}_c=R^a{}_c+3\delta^a_c,
\qquad
\mathfrak R^a=R^a{}_c e^c+3e^a.
\label{eq:AdS-Ricci-shift-main}
\end{equation}
We shall also use the differential forms
\begin{align}
\mathfrak B^a&:=\hat R^{ab}\wedge e_b,
&
\mathfrak C&:=e_a\wedge\mathfrak R^a,
\\
\mathfrak A&:=e_a\wedge T^a,
&
\mathfrak t&:=\iota_aT^a.
\label{eq:derived-gravitational-forms-main}
\end{align}
Here $\mathfrak B^a$ is the curvature Bianchi three-form, $\mathfrak C$ is a two-form that measures the antisymmetric part of the Ricci tensor, and $\mathfrak A$ and $\mathfrak t$ are the axial three-form and trace one-form of the torsion, respectively. The trace one-form of the torsion can also be written as
\begin{align}
\mathfrak t
:=\iota_a T^a
&=
\frac12 T^a{}_{bc}\,
\iota_a\!\left(e^b\wedge e^c\right)
\nonumber\\
&=
\frac12 T^a{}_{bc}
\left(\delta_a^b e^c-\delta_a^c e^b\right)
\nonumber\\
&=
T^a{}_{ab}\,e^b.
\label{eq:torsion-trace-one-form}
\end{align}

In terms of these forms, the gravitational restriction of each invariant in
Eq.~\eqref{LLlagrangian} is
\begin{align}
\left.\mathcal I_\alpha\right|_{\mathrm{grav}}
={}&-\frac12\hat R^{ab}\wedge*\hat R_{ab}
+T^a\wedge*T_a,
\label{eq:Ialpha-form-main}
\\[2pt]
\left.\mathcal I_\beta\right|_{\mathrm{grav}}
={}&-\mathfrak B^a\wedge*\mathfrak B_a
-\mathfrak R^a\wedge*\mathfrak R_a
+T^a\wedge*T_a
+\mathfrak A\wedge*\mathfrak A
+\mathfrak t\wedge*\mathfrak t,
\label{eq:Ibeta-form-main}
\\[2pt]
\left.\mathcal I_\gamma\right|_{\mathrm{grav}}
={}&\epsilon_{abcd}\hat R^{ab}\wedge\hat R^{cd}
-2\hat R^{ab}\wedge*\hat R_{ab}
+4\mathfrak B^a\wedge*\mathfrak B_a
+4\mathfrak R^a\wedge*\mathfrak R_a
\nonumber\\
&-4\mathfrak A\wedge*\mathfrak A
-4\mathfrak t\wedge*\mathfrak t,
\label{eq:Igamma-form-main}
\\[2pt]
\left.\mathcal I_\delta\right|_{\mathrm{grav}}
={}&-4\mathfrak B^a\wedge*\mathfrak B_a
+4T^a\wedge*T_a
+4\mathfrak t\wedge*\mathfrak t,
\label{eq:Idelta-form-main}
\\[2pt]
\left.\mathcal I_\varepsilon\right|_{\mathrm{grav}}
={}&16\mathfrak C\wedge*\mathfrak C
-16\mathfrak A\wedge*\mathfrak A,
\label{eq:Iepsilon-form-main}
\\[2pt]
\mathcal I_\zeta
={}&16\vol.
\label{eq:Izeta-form-main}
\end{align}
Thus the complete purely gravitational four-form, including the algebraic
volume contribution, can be collected as
\begin{align}
L_{\mathrm{grav}}
={}&\gamma\epsilon_{abcd}\hat R^{ab}\wedge\hat R^{cd}
-\left(\frac{\alpha}{2}+2\gamma\right)
 \hat R^{ab}\wedge*\hat R_{ab}
\nonumber\\
&+\left(-\beta+4\gamma-4\delta\right)
 \mathfrak B^a\wedge*\mathfrak B_a
+\left(-\beta+4\gamma\right)
 \mathfrak R^a\wedge*\mathfrak R_a
\nonumber\\
&+\left(\alpha+\beta+4\delta\right)T^a\wedge*T_a
+\left(\beta-4\gamma-16\varepsilon\right)
 \mathfrak A\wedge*\mathfrak A
\nonumber\\
&+\left(\beta-4\gamma+4\delta\right)
 \mathfrak t\wedge*\mathfrak t
+16\varepsilon\mathfrak C\wedge*\mathfrak C
+16\zeta\vol.
\label{eq:Lgrav-full-form-main}
\end{align}

This form of the gravity action contains no unevaluated Clifford traces. Its detailed derivation and its conversion to the tensor-component basis used in the spectral comparison are presented in Appendix~\ref{app:table1-derivation}. In particular, the coefficient choice~\eqref{eq:torsional-ads-gravity} reduces the gravity sector to the Lorentz-projected AdS Euler form. The resulting theory is therefore a Riemann--Cartan extension of the MacDowell--Mansouri formulation of AdS gravity. In the torsionless limit, it reduces, up to the topological Euler term, to ordinary Einstein gravity with a negative cosmological constant.

\subsection{Fermionic Sector}\label{sec:fermion-form}

We next isolate the terms that are quadratic in the fermions and contain no
explicit curvature, torsion, or internal field strength.  According to
Eq.~\eqref{Eq_X}, the odd component of the supercurvature can be decomposed as
\begin{equation}
X_i=X_i^{(0)}+X_i^{(T)},
\qquad
X_i^{(0)}:=-\se\wedge\mathrm D\psi_i
+\frac12\se^2\psi_i,
\qquad
X_i^{(T)}:=T^a\gamma_a\psi_i.
\label{eq:Xsplit-main}
\end{equation}
The contribution $X_i^{(T)}$, as well as the fermion-dependent corrections to
the bosonic curvature components, generates nonminimal or higher-order
interactions and is included in $L_{\mathrm{NMC}}$.  It will not be included in
the present subsection.  The purely fermionic curvature considered here is
therefore
\begin{equation}
\left.\FF\right|_{\mathrm F}^{(0)}
=\overline{\boldsymbol Q}{}^i_\alpha
\left(X_i^{(0)}\right)^\alpha.
\label{eq:pure-fermionic-curvature-main}
\end{equation}

Let $\bar X^{(0)i}$ denote the Majorana conjugate of $X_i^{(0)}$.  All the
purely fermionic projections of the curvature-dependent invariants can be
written in terms of the following two differential-form contractions:
\begin{align}
\mathcal A_X^{(0)}
&:=\bar X^{(0)i}\wedge*X_i^{(0)},
\label{eq:AX0-main}
\\
\mathcal B_X^{(0)}
&:=\bar X^{(0)i}\gamma_{ab}\wedge e^a
\wedge*\!\left(X_i^{(0)}\wedge e^b\right).
\label{eq:BX0-main}
\end{align}
The internal index $i$ is summed.  Direct multiplication of the odd
supermatrix blocks, using the ordered supertrace and
$\langle\overline{\boldsymbol Q}{}^i_\alpha
\overline{\boldsymbol Q}{}^j_\beta\rangle
=-2C_{\alpha\beta}\delta^{ij}$, gives
\begin{align}
\left.\mathcal I_\alpha\right|_{\mathrm F}^{(0)}
&=2\mathcal A_X^{(0)},
&
\left.\mathcal I_\beta\right|_{\mathrm F}^{(0)}
&=2\mathcal A_X^{(0)}+\mathcal B_X^{(0)},
\nonumber\\
\left.\mathcal I_\gamma\right|_{\mathrm F}^{(0)}
&=-6\mathcal B_X^{(0)},
&
\left.\mathcal I_\delta\right|_{\mathrm F}^{(0)}
&=4\mathcal A_X^{(0)}+2\mathcal B_X^{(0)},
\nonumber\\
\left.\mathcal I_\varepsilon\right|_{\mathrm F}^{(0)}
&=-12\mathcal B_X^{(0)},
&
\left.\mathcal I_\zeta\right|_{\mathrm F}^{(0)}
&=0.
\label{eq:fermionic-invariant-map-main}
\end{align}
It follows that the purely fermionic part of the master Lagrangian is the
four-form
\begin{align}
L_{\mathrm{fer}}
={}&2\left(\alpha+\beta+2\delta\right)
\mathcal A_X^{(0)}
\nonumber\\
&+\left(\beta-6\gamma+2\delta-12\varepsilon\right)
\mathcal B_X^{(0)}.
\label{eq:Lfer-differential-form-main}
\end{align}
Equivalently, inserting the definitions~\eqref{eq:AX0-main} and
\eqref{eq:BX0-main},
\begin{align}
L_{\mathrm{fer}}
={}&2\left(\alpha+\beta+2\delta\right)
\bar X^{(0)i}\wedge*X_i^{(0)}
\nonumber\\
&+\left(\beta-6\gamma+2\delta-12\varepsilon\right)
\bar X^{(0)i}\gamma_{ab}\wedge e^a
\wedge*\!\left(X_i^{(0)}\wedge e^b\right).
\label{eq:Lfer-expanded-differential-form-main}
\end{align}
It is important to emphasize that neither $\mathcal A_X^{(0)}$ nor
$\mathcal B_X^{(0)}$ has a definite derivative order.  Since $X_i^{(0)}$
contains both $\mathrm D\psi_i$ and $\psi_i$, products of two derivative
terms, mixed products, and purely algebraic products respectively generate
second-order, first-order, and mass-like fermionic terms.  To display this
separation explicitly, define
\begin{align}
\mathcal K_D
&:=\bar\psi^i\gamma^\mu\mathrm D_\mu\psi_i
-\mathrm D_\mu\bar\psi^i\gamma^\mu\psi_i,
&
\mathcal M&:=\bar\psi^i\psi_i,
\nonumber\\
\mathcal K_2
&:=\mathrm D^\mu\bar\psi^i\mathrm D_\mu\psi_i,
&
\mathcal K_\gamma
&:=\mathrm D_\mu\bar\psi^i\gamma^{\mu\nu}
\mathrm D_\nu\psi_i.
\label{eq:fermionic-component-structures-main}
\end{align}
The two geometric contractions then have the distinct component expansions
\begin{align}
\mathcal A_X^{(0)}
&=\vol\left(
3\mathcal K_D-3\mathcal M+3\mathcal K_2+\mathcal K_\gamma
\right),
\label{eq:AX0-component-expansion-main}
\\
\mathcal B_X^{(0)}
&=\vol\left(
12\mathcal K_D-12\mathcal M+6\mathcal K_2+6\mathcal K_\gamma
\right).
\label{eq:BX0-component-expansion-main}
\end{align}
For clarity, introduce
\begin{equation}
a_F:=2\left(\alpha+\beta+2\delta\right),
\qquad
b_F:=\beta-6\gamma+2\delta-12\varepsilon,
\label{eq:fermionic-form-coefficients-main}
\end{equation}
so that $L_{\mathrm{fer}}=a_F\mathcal A_X^{(0)}
+b_F\mathcal B_X^{(0)}$.  Equations~\eqref{eq:AX0-component-expansion-main}
and~\eqref{eq:BX0-component-expansion-main} give
\begin{align}
\frac{L_{\mathrm{fer}}}{\vol}
={}&\left(3a_F+12b_F\right)\mathcal K_D
-\left(3a_F+12b_F\right)\mathcal M
\nonumber\\
&+\left(3a_F+6b_F\right)\mathcal K_2
+\left(a_F+6b_F\right)\mathcal K_\gamma.
\label{eq:Lfer-component-weights-main}
\end{align}
Consequently, the four coefficients are
\begin{align}
C_D
&=6\alpha+18\beta-72\gamma+36\delta-144\varepsilon,
&
C_M&=-C_D,
\nonumber\\
C_2
&=6\alpha+12\beta-36\gamma+24\delta-72\varepsilon,
&
C_\gamma
&=2\alpha+8\beta-36\gamma+16\delta-72\varepsilon.
\label{eq:fermionic-component-coefficients-main}
\end{align}
In particular, the first-order Dirac coefficient $C_D$ and the
Klein--Gordon-type coefficient $C_2$ are different.  The latter factorizes as
\begin{equation}
C_2
=6\left(
\alpha+2\beta-6\gamma+4\delta-12\varepsilon
\right),
\label{eq:KG-coefficient-factorization-main}
\end{equation}
so its cancellation is precisely the condition~\eqref{Eq_Coef_KleinGordon}.
The expression above contains the kinetic, mass-like, and purely fermionic
two-derivative structures generated by the composite odd connection
$\se\psi_i$, but no terms assigned to $L_{\mathrm{NMC}}$.  The detailed
Clifford reduction is given in Appendix~\ref{app:table1-derivation}.

\section{Ghost-Free Gravity Sector}\label{sec:ghost-free}

The explicit reduction of the Clifford-generated invariants provides the coefficient map relating the geometric action~\eqref{LLlagrangian} to the usual curvature-, torsion-, gauge-, and fermion-dependent terms. Besides establishing the component content of the unified action, these results supply valuable information for analyzing its propagating degrees of freedom and those of models obtained through consistent restrictions or truncations. In this section, we use this coefficient map to determine whether the reduced bosonic gravitational sector intersects the parameter regions previously identified as ghost free at the linearized level. This comparison is intended as a nontrivial consistency check and as a starting point for subsequent spectral analyses; we do not claim that it establishes ghost freedom of the complete theory, whose coupled fermionic, gauge, nonminimal, and nonlinear sectors require a separate constraint and stability analysis.

The six coefficients in the Lagrangian~\eqref{LLlagrangian} are not fixed by a single requirement, and it is useful to distinguish constraints that follow directly from the unified theory from those obtained by comparison with a reduced bosonic model. The sign of the internal Yang--Mills sector fixes $\alpha < 0$, while the absence of the Klein--Gordon--type fermionic kinetic term imposes the independent relation~\eqref{Eq_Coef_KleinGordon}.  The coefficient $\zeta$ does not enter either of these conditions; it controls the purely algebraic frame term and hence the cosmological contribution.  Thus, already at this stage, the unified action contains nontrivial relations among its gauge, fermionic, and gravitational sectors without requiring a ghost analysis of the complete coupled system.

A complete ghost analysis of $L_{\mathrm{LL}}$, including the fermions, the internal gauge fields, and the nonminimal couplings generated by the full supercurvature, is substantially more involved and lies beyond the scope of this work.  However, one may truncate to the purely bosonic gravitational sector by setting the fermions to zero, suppressing internal gauge-field fluctuations and choosing the $\zeta$-coefficient so as to have vanishing cosmological constant. The resulting theory is a parity--even quadratic Poincar\'e gauge model, equivalent to the reduced bosonic theory defined by restricting the superalgebra-valued action to its spacetime gauge sector. Around a Minkowski background, its linearized spectrum can be compared directly with the analysis of Refs.~\cite{Sezgin:1979zf,Sezgin:1981xs}.  The corresponding ghost--free conditions then impose additional relations among the coefficients of the reduced bosonic action.  These relations should not be interpreted as a proof of ghost freedom of the complete theory; rather, they test whether the Clifford--generated coefficient map intersects the known healthy regions of the linearized bosonic theory.

To make this comparison explicit, Table~\ref{tab:SvN-ModGB} lists the coefficients of the tensorial invariants generated by Eq.~\eqref{LLlagrangian}.  The algebraic reduction required to obtain these coefficients is substantial, involving the evaluation of the relevant superalgebra traces, Clifford-algebra identities, and the subsequent decomposition into independent tensorial invariants; the main steps of this calculation are summarized in Appendix~\ref{app:table1-derivation}. For direct comparison with Ref.~\cite{Sezgin:1981xs}, the $R^2$ term has been eliminated using the four-dimensional Euler (Gauss--Bonnet) density, i.e. modulo the corresponding topological contribution.  The same table also records the gauge and purely fermionic quadratic terms, so that the origin of the normalization condition \eqref{Eq_alpha_1/2_by_YM} and of the fermionic constraint~\eqref{Eq_Coef_KleinGordon} is manifest.

\begin{table}[t]
	\centering
	\begin{tabular}{cc}
		\hline
		Invariant term & Coefficient \\
		\hline \\[-6pt]
		$R$ &  $ -\alpha - 6\beta + 24\gamma$ \\[2pt]
		
		Constant volume term $C_0$ & $-6\alpha - 36\beta + 144\gamma + 16\zeta$ \\[2pt]
		
		$R_{abcd} R^{abcd}$ & $-\alpha/4 - \beta/2 + \gamma - 2\delta$ \\[2pt]
		
		$R_{abcd} R^{acbd}$ & $\beta - 4\gamma + 4\delta$ \\[2pt]
		
		$R_{abcd} R^{cdab}$ & 0 \\[2pt]
		
		$R_{ab} R^{ab}$ & $-\beta + 4\gamma + 16\varepsilon$ \\[2pt]
		
		$R_{ab} R^{ba}$ & $-16\varepsilon$\\[2pt]
		
		$T_{abc}\,T^{abc}$ & $\alpha/2 + \beta - 2\gamma + 2\delta - 8\varepsilon$ \\[2pt]
		
		$T_{abc}\,T^{bca}$ & $\beta - 4\gamma - 16\varepsilon$ \\[2pt]
		
		$T_{b}{}^{ba}\,T^{c}{}_{ca}$ & $\beta - 4\gamma + 4\delta$ \\[2pt]
		
		$f^{ij}{}_{\,ab} f_{ij}{}^{ab}$ & $\alpha/2$\\[2pt]
		
		$\bar{\psi}\gamma^{\mu}\mathrm{D}_{\mu}\psi
		-\mathrm{D}_{\mu}\bar{\psi}\gamma^{\mu}\psi$ & $6 \alpha + 18 \beta - 72 \gamma + 36 \delta - 144 \varepsilon$\\[2pt]
		
		$\bar{\psi}\psi$ & $(-1)\times$ row above\\[2pt]
		
		$\mathrm{D}^{\mu}\bar{\psi}\,\mathrm{D}_{\mu}\psi$ & $6 \alpha + 12 \beta - 36 \gamma + 24 \delta - 72 \varepsilon$\\[2pt]
		
		$\mathrm{D}_{\mu}\bar{\psi}\, \gamma^{\mu\nu} \, \mathrm{D}_{\nu}\psi$ & $2 \alpha + 8 \beta - 36 \gamma + 16 \delta - 72 \varepsilon$\\[2pt]
		\hline
	\end{tabular}
	\caption{Coefficients of curvature invariants obtained from Eq.~\eqref{LLlagrangian}. All expressions are written in terms of tensor components. The Gauss--Bonnet theorem has been used to eliminate $R^{2}$ in favor of quadratic contractions of $R_{ab}$ and $R_{abcd}$. When the internal gauge coupling is set to unity and the gauge fields are canonically normalized, $\alpha$ takes the value given in \eqref{Eq_alpha_1/2_by_YM}.}
	\label{tab:SvN-ModGB}
\end{table}

We next exhibit three representative algebraic branches obtained by matching the coefficient map of the reduced bosonic theory to the parameter relations of the models containing a massless $\mathbf{2}^{+}$ mode identified in Ref.~\cite{Sezgin:1981xs}. As shown below, positivity of the Einstein--Hilbert and Yang--Mills kinetic terms, at the level of the reduced bosonic theory, retains the three models. Let us remark that Table~\ref{tab:modelsSvZ} should therefore be read primarily as a consistency check of the coefficient map and of its intersection with the linearized bosonic classification. It is not a statement that every displayed branch defines a physically admissible sector of the unified theory.

The physical interpretation of the propagating $\mathbf{2}^{+}$, $\mathbf{1}^{\pm}$, and $\mathbf{0}^{\pm}$ modes appearing in Table~\ref{tab:modelsSvZ} is reviewed in Appendix~\ref{app:linearizedtorsionmodes}.  The standard fermionic first-order term $\bar{\psi}\gamma^{\mu}\mathrm{D}_{\mu}\psi -\mathrm{D}_{\mu}\bar{\psi}\gamma^{\mu}\psi$ originates from cross terms in Eq.~\eqref{Eq_X}, and a field renormalization is required to bring it to the canonical Majorana normalization. 

\begin{table}[h]
	\centering
	\begin{tabular}{ p{2.7cm} p{1.4cm} p{1.7cm} p{2.4cm} p{0.4cm}}
		\hline
		\makecell{Algebraic branch\\and defining\\conditions} & $C_R$ & GPC & \makecell{Additional\\ condition\\for KG($\psi$) = 0} & ST \\ \hline
		``Model 1'' & \multirow{3}{*}{$-\alpha -96\varepsilon$}  & \multirow{3}{*}{$\mathbf{2}^{+},\mathbf{0}^{+},\mathbf{0}^{-}$} &  \multirow{3}{*}{$\varepsilon\!\to\! \beta/24$} & \multirow{3}{*}{\checkmark} \\
		$\gamma\!\to\! \beta/4 - 4\varepsilon$ &&&&\\
		$\delta\!\to\!-(\alpha+\beta)/4$ &&&&\\ \hline

		``Model 3''  & \multirow{4}{*}{$-\alpha - 96\varepsilon$} & \multirow{4}{*}{$\mathbf{2}^{+},\mathbf{1}^{-},\mathbf{0}^{-}$} & \multirow{4}{*}{$\varepsilon\!\to\!-\alpha/72$} & \multirow{4}{*}{\checkmark} \\
		$\beta\!\to\!-\alpha -48\varepsilon$ &&&& \\
		$\gamma\!\to\!-(\alpha+64\varepsilon)/4$ &&&& \\
		$\delta\!\to\! 12\varepsilon$ &&&& \\ \hline

		``Model 5''  & \multirow{4}{*}{$-\alpha$} & \multirow{4}{*}{$\mathbf{2}^{+},\mathbf{1}^{+},\mathbf{0}^{-}$} & \multirow{4}{*}{$\varepsilon\!\to\!\alpha/24$} & \multirow{4}{*}{\xmark}\\
		 $\beta\!\to\!-\alpha$ &&&&\\
		 $\gamma\!\to\!-\alpha/4$ &&&&\\
		$\delta\!\to\!0$ &&&&\\ \hline
	\end{tabular}
	\caption{Algebraic branches obtained by matching the reduced bosonic coefficient map of Eq.~\eqref{LLlagrangian} and Table~\ref{tab:SvN-ModGB} to the parameter relations identified in the linearized Minkowski analysis of Ref.~\cite{Sezgin:1981xs}.
	The first column lists the parameter relations defining each reduced bosonic model.
	Here GPC denotes the gravitational particle content, labeled by irreducible representations of $so(3)$.
	The fourth column gives the additional restrictions obtained by intersecting the parameter relations in the first column with the codimension-one condition~\eqref{Eq_Coef_KleinGordon}.
	ST indicates whether the first-order spinor kinetic term survives.
	The displayed value of $C_R$ is evaluated after imposing the bosonic parameter relations of the first column.
	}
	\label{tab:modelsSvZ}
\end{table}

The comparison with Refs.~\cite{Sezgin:1979zf,Sezgin:1981xs} is performed around flat spacetime and therefore requires the coefficient of the constant volume term to vanish. In the present theory, the coefficients of the Ricci scalar and constant terms are, respectively,
\begin{equation}
C_R=-\alpha-6\beta+24\gamma,
\qquad
C_0=-6\alpha-36\beta+144\gamma+16\zeta.
\end{equation}
Upon matching
\begin{equation}
C_R R+C_0
=
\frac{1}{16\pi G_{\mathrm N}}(R-2\Lambda),
\end{equation}
one obtains
\begin{equation}
C_R=\frac{1}{16\pi G_{\mathrm N}},
\qquad
\Lambda=-\frac{C_0}{2C_R}
       =-8\pi G_{\mathrm N}C_0.
\end{equation}
Consequently,
\begin{equation}
\Lambda
=
-\frac{3\alpha+18\beta-72\gamma-8\zeta}
{\alpha+6\beta-24\gamma}.
\end{equation}
For each algebraic branch displayed in Table~\ref{tab:modelsSvZ}, the coefficient $\zeta$ may be chosen so that $C_0=0$. This establishes the existence of a Minkowski background at the level of the constant term, but does not alter the positivity conditions on $C_R$. Within the reduced bosonic sector, positivity of the Einstein--Hilbert term is compatible with the positivity of the Yang--Mills term in the three branches given by Model~1, Model~3 and Model~5. For Model~5, the additional condition $\varepsilon=\alpha/24$ causes all purely fermionic quadratic terms displayed in Table~\ref{tab:SvN-ModGB} to vanish. In particular, both the Klein--Gordon--type and first-order Dirac kinetic terms cancel. This does not imply that the complete theory is purely bosonic, since nonminimal and higher-order fermionic interactions may remain. This Model~5 subbranch is consistent with the action \eqref{AdSaction}.

Model~3 becomes inadmissible once the spinor sector is included: imposing condition~\eqref{Eq_Coef_KleinGordon} gives $C_R=\alpha/3<0$ and hence the wrong sign for the Einstein--Hilbert term.
For Model~1, the condition \eqref{Eq_Coef_KleinGordon} does not contradict the positivity of $C_R$ in the region $\beta<-\alpha/4$.

Even within the purely bosonic Poincar\'e gauge sector, linearized ghost freedom does not by itself establish nonlinear stability. Propagating vector and axial--vector torsion modes can acquire additional degrees of freedom or mix with the metric sector once nonlinear interactions are included \cite{BeltranJimenez:2019hrm}.  Conversely, special degeneracies or additional geometric mechanisms may evade some of these problems \cite{Barker:2022jsh,Barker:2024goa}.  

The distinction is still more important for the complete unified theory. Because the bosonic supercurvatures themselves contain fermion bilinears, as displayed explicitly around Eq.~\eqref{Eq_Lorentz_Curvature}, and because $L_{\mathrm{NMC}}$ couples fermions to curvature, torsion, and the internal gauge field strength, the constraint structure of the coupled system need not coincide with that of its bosonic truncation.  Establishing ghost freedom of the full theory would require an analysis of the complete coupled kinetic and constraint system, including the nonminimal interactions and, in general, backgrounds beyond Minkowski space.  We leave this problem for future work.

\section{Discussion and Outlook}

In this work we have shown that a single geometric action, constructed as a linear combination of a restricted family of Clifford-generated invariants, simultaneously produces Yang--Mills, fermionic, and quadratic gravitational sectors with torsion. For specific algebraic relations among its coefficients, the reduced bosonic sector reproduces known linearized ghost--free models with propagating torsion. After canonical normalization of the Yang--Mills sector, all three branches can intersect the region $C_R>0$ at the level of the reduced bosonic theory. After additionally imposing the absence of the Klein--Gordon--type fermionic kinetic term, Model~3 is excluded because $C_R=\alpha/3<0$, whereas Model~5 remains admissible and Model~1 requires $\beta<-\alpha/4$.

The fermionic sector provides an additional consistency check. Although the composite connection component $\slashed e\psi$ generically produces both first- and second--order derivative structures, the independent Klein--Gordon--type term vanishes on the locus \eqref{Eq_Coef_KleinGordon}. The remaining antisymmetric two--derivative term can be rewritten, up to boundary contributions, as nonminimal curvature and torsion couplings. Thus, on this parameter locus, the independent propagating fermionic kinetic structure is generically the standard first-order Dirac term. The Model~5 intersection is exceptional, since its purely fermionic quadratic derivative terms cancel simultaneously.

Given the presence of quadratic curvature terms, the resulting theory may be interpreted as an effective field theory below the characteristic scale of the additional torsional modes. Within such an interpretation, higher--order invariants would be organized according to the corresponding effective field theory expansion. Establishing the associated cutoff and power counting requires a separate analysis.

Several directions for further investigation naturally follow. First is the question of identifying representations capable of accommodating the Standard Model gauge group and fermion content. Also, a complete ghost stability analysis, including the effects of nonminimal couplings on more general backgrounds, is clearly required. Other promising avenues include the extension to superconformal algebras accommodating chiral fermions, the study of dynamical mechanisms that select the Lorentz subgroup, and the exploration of cosmological and phenomenological implications of massive torsion modes and nonminimal couplings. In particular, the possibility of dynamical dark energy contributions and novel dark matter induced by torsion and nonminimal couplings appears especially intriguing.

More broadly, the results presented here suggest that Clifford--algebra--based constructions offer a natural, economical, and conceptually transparent arena for unifying gravity and gauge interactions. They do so without invoking extra dimensions or enlarging the field content beyond that already implied by geometry itself, pointing toward a unifying purely geometric principle.

\section*{Acknowledgements}
P.A. acknowledges support from ANID-FONDECYT Regular grant No. 1230112.
C.Q. acknowledges support from ANID-FONDECYT grant 11231238.
P.A. and F.I. dedicate this work to their daughters Lizzie and Luc\'ia, whose lovely antics and cheerful laugh made the long algebraic calculations and coding time bearable. The geometric Lagrangian (\ref{LLlagrangian}) for all the interactions is named after them.
F.I. is also grateful for the emotional support provided by the Netherlands Bach Society. They made freely available superb quality recordings of the music of Bach, and without them, this work would have been impossible.

\begin{appendices}

\section{Useful definitions and properties}

The tangent-space gamma matrices satisfy
\begin{equation}
 \mathrm{d}\gamma^a = 0 \,, \qquad \mathrm{D}\gamma^a = 0 \, ,
\end{equation}
where the Lorentz-covariant derivative acts as
\begin{equation}
 \mathrm{D}\gamma^a\equiv \mathrm{d}\gamma^a +\omega^a{}_b \gamma^b + \frac{1}{2}\omega^{bc}\Sigma_{bc}\gamma^a - \frac{1}{2}\omega^{bc}\gamma^a\Sigma_{bc} \, .
\end{equation}

Optionally, as invariant algebra tensors, one may use the fully symmetrized supertrace of $n$ homogeneous superalgebra-valued elements $X_1,\dots,X_n$. It is defined by
\begin{equation}
\label{symstr}
\langle X_1\cdots X_n\rangle_{\mathrm{FS}}
\equiv
\frac{1}{n!}
\sum_{\pi\in S_n}
(-1)^{\varepsilon(\pi;X)}
\left\langle
X_{\pi(1)}X_{\pi(2)}\cdots X_{\pi(n)}
\right\rangle,
\end{equation}
where $S_n$ is the permutation group of $n$ elements and $(-1)^{\varepsilon(\pi;X)}$ is the Koszul sign generated by permuting the homogeneous elements $X_i$. Explicitly,
\begin{equation}
\label{koszul-sign}
(-1)^{\varepsilon(\pi;X)}
=
\prod_{\substack{r<s\\ \pi(r)>\pi(s)}}
(-1)^{|X_{\pi(r)}|\,|X_{\pi(s)}|},
\end{equation}
where $|X_i|=0$ for bosonic (even) elements and $|X_i|=1$ for fermionic (odd) elements. Equivalently, each interchange of two homogeneous elements $X_i$ and $X_j$ contributes the factor $(-1)^{|X_i||X_j|}$.

The fully symmetrized supertrace can be useful for simplifying algebraic manipulations. However, it is not used in deriving the results displayed in Table~\ref{tab:SvN-ModGB}; there we use standard ordered (unsymmetrized) supertraces, as detailed in Appendix~\ref{app:table1-derivation}.

For the form algebra we use the orientation
\begin{equation}
\epsilon_{0123}=+1,
\qquad
\epsilon^{0123}=-1,
\end{equation}
where indices on the Levi--Civita tensor are raised with $\eta_{ab}=\operatorname{diag}(-,+,+,+)$. The Hodge dual acts on the orthonormal basis as
\begin{equation}
*
\left(
e^{a_1}\wedge\cdots\wedge e^{a_p}
\right)
=
\frac{1}{(4-p)!}\,
\epsilon^{a_1\cdots a_p}{}_{b_{p+1}\cdots b_4}\,
e^{b_{p+1}}\wedge\cdots\wedge e^{b_4},
\qquad
p=0,\ldots,4.
\end{equation}
therefore
\begin{equation}
e^{a_1}\wedge\cdots\wedge e^{a_p}\wedge *\bigl(e^{b_1}\wedge\cdots\wedge e^{b_p}\bigr)
= p! \delta^{a_1\cdots a_p}_{c_1\cdots c_p} \eta^{c_1b_1}\cdots\eta^{c_pb_p}\,\vol,
\label{eq:app-hodge-basis-paper}
\end{equation}
where
\begin{equation}
\delta^{a_1\cdots a_p}_{c_1\cdots c_p}
:=
\delta^{[a_1}_{c_1}\cdots
\delta^{a_p]}_{c_p}
=
\frac{1}{p!}
\sum_{\sigma\in S_p}
\operatorname{sgn}(\sigma)\,
\delta^{a_1}_{c_{\sigma(1)}}\cdots
\delta^{a_p}_{c_{\sigma(p)}}
=
\frac{1}{p!}
\det\!\left(\delta^{a_i}_{c_j}\right).
\end{equation}
With $\vol=\sqrt{|g|}\,d^4x = e^0\wedge e^1\wedge e^2\wedge e^3 $. For $p=1$ we get
\begin{equation}
 e^a \wedge *\bigl(e^b\bigr) = \eta^{ab} \vol\,.
\end{equation}

\section{Explicit reduction of the Clifford--superalgebra invariants}
\label{app:table1-derivation}

This appendix gives the component reduction of the six invariants entering Eq.~\eqref{LLlagrangian}.  In particular, the superconnection and its curvature are denoted by $\AAA$ and $\FF$, their components by $F^{ab}$, $F^a$, $F^{ij}$ and $X_i^\alpha$. The calculation separates naturally into the Clifford-algebra part, evaluated in the $4\times4$ spacetime block of the $osp(n|4)$ supermatrix representation, and the exterior-form part,
evaluated with the ordinary spacetime Hodge dual.

\subsection{Supermatrix realization and Clifford conventions}

We use the block realization introduced in Eq.~\eqref{superalgebrarep}.  For an even block-diagonal supermatrix
\begin{equation}
\mathbb M=
\left[
\begin{array}{c|c}
M_{4\times4}&0\\ \hline
0&N_{n\times n}
\end{array}
\right],
\end{equation}
The ordered (unsymmetrized) super trace is defined as regular traces in the block-diagonal matrices
\begin{align}
\langle \mathbb M \rangle &= \operatorname{STr}\mathbb M\\
&=\operatorname{Tr}_{4}M-\operatorname{Tr}_{n}N.
\end{align}
The Clifford generators are embedded as in the main text,
\begin{equation}
\Gamma_{a_1 \cdots a_p}=
\left[
\begin{array}{c|c}
\gamma_{a_1 \cdots a_p}&0\\ \hline
0&0
\end{array}
\right],
\end{equation}
where
\begin{equation}
\{\gamma_a,\gamma_b\}=2\eta_{ab},
\qquad
\gamma_{ab}:=\frac12[\gamma_a,\gamma_b],
\qquad
\eta_{ab}=\operatorname{diag}(-,+,+,+).
\end{equation}

The spacetime generators have support only of the upper block,
\begin{equation}
\boldsymbol P_a=\frac12\Gamma_a,
\qquad
\boldsymbol\Sigma_{ab}
=\frac14[\Gamma_a,\Gamma_b]
=\frac12
\left[
\begin{array}{c|c}
\gamma_{ab}&0\\ \hline
0&0
\end{array}
\right],
\label{eq:app-block-generators}
\end{equation}
and the internal generators have support only on the lower block,
\begin{equation}
\boldsymbol L_{ij}=
\left[
\begin{array}{c|c}
0&0\\ \hline
0&(L_{ij})^k{}_l
\end{array}
\right],
\qquad
(L_{ij})^k{}_l
=\delta_i^k\delta_{jl}-\delta_j^k\delta_{il}.
\end{equation}
Accordingly,
\begin{equation}
\se=e^a\Gamma_a=2e^a\boldsymbol P_a,
\qquad
\se^2=e^a\wedge e^b\,\Gamma_a\Gamma_b
=2e^a\wedge e^b\,\boldsymbol\Sigma_{ab}.
\label{eq:app-se-square-paper}
\end{equation}

For completeness, we also give the odd generators explicitly.  Let $\mathbf e_\alpha$ ($\alpha=1,\ldots,4$) and $\mathbf u_i$ ($i=1,\ldots,n$) denote the standard basis column vectors of the spinor and internal spaces, respectively, and let $C_{\alpha\beta}$ be the charge conjugation matrix.  We use
\begin{equation}
C^T=-C,
\qquad
C_{\alpha\gamma}C^{\gamma\beta}=\delta_\alpha{}^\beta,
\qquad
C\gamma_a C^{-1}=-(\gamma_a)^T.
\label{eq:app-C-properties}
\end{equation}
Defining the spinor column $\mathbf c^{\alpha}:=C^{\alpha\beta}\mathbf e_\beta$, the two equivalent index placements of the Majorana supercharges are represented by the $(4+n)\times(4+n)$ odd supermatrices
\begin{align}
\boldsymbol Q_i^{\alpha}
&=
\left[
\begin{array}{c|c}
0_{4\times4} & \mathbf c^{\alpha}\mathbf u_i^{T} \\\hline
\mathbf u_i\mathbf e_\alpha^{T} & 0_{n\times n}
\end{array}
\right],
\label{eq:app-Q-matrix}
\\[0.5em]
\overline{\boldsymbol Q}{}^{i}_{\alpha}
&=
\left[
\begin{array}{c|c}
0_{4\times4} & \mathbf e_\alpha\mathbf u_i^{T} \\\hline
\mathbf u_i\mathbf e_\alpha^{T}C & 0_{n\times n}
\end{array}
\right].
\label{eq:app-Qbar-matrix}
\end{align}
The internal metric is $\delta_{ij}$, and the Majorana reality condition is implemented algebraically as
\begin{equation}
\overline{\boldsymbol Q}{}^{i}_{\alpha}
=C_{\alpha\beta}\,\delta^{ij}\boldsymbol Q_j^{\beta}.
\label{eq:app-Q-Majorana-relation}
\end{equation}
Equivalently, if $A,B=1,\ldots,4+n$ denote full supermatrix indices, the nonzero entries are
\begin{align}
(\overline{\boldsymbol Q}{}^i_\alpha)^\beta{}_{4+j}
&=\delta^\beta_\alpha\delta^i_j,
&
(\overline{\boldsymbol Q}{}^i_\alpha)^{4+j}{}_{\beta}
&=\delta^j_i C_{\alpha\beta},
\label{eq:app-Qbar-components}
\\
(\boldsymbol Q_i^\alpha)^\beta{}_{4+j}
&=C^{\alpha\beta}\delta_{ij},
&
(\boldsymbol Q_i^\alpha)^{4+j}{}_{\beta}
&=\delta_i^j\delta^\alpha_\beta.
\label{eq:app-Q-components}
\end{align}
These are precisely the off--diagonal blocks indicated schematically by $Q_{4\times n}$ and $Q^T_{n\times4}$ in Eq.~\eqref{superalgebrarep}.  Direct matrix multiplication gives
\begin{equation}
\operatorname{STr}\!\left(
\boldsymbol Q_i^\rho\overline{\boldsymbol Q}{}^j_\sigma
\right)
=\delta_i^j
\left(C_{\sigma\beta}C^{\rho\beta}-\delta^\rho_\sigma\right)
=-2\,\delta_i^j\delta^\rho_\sigma,
\label{eq:app-QbarQ-direct}
\end{equation}
where $C_{\sigma\beta}C^{\rho\beta}=-\delta^\rho_\sigma$ follows from the antisymmetry of $C$.  This reproduces the fermionic quadratic invariant used below and, using the Clifford identities in the upper block, the same matrix representation reproduces the supercharge anticommutator \eqref{eq:app-supercharge-anticommutator}. The quadratic invariant tensors needed below are therefore
\begin{align}
\langle \boldsymbol P_a\boldsymbol P_b\rangle
&=\eta_{ab},
\label{eq:app-PP}
\\
\langle \boldsymbol\Sigma_{ab}\boldsymbol\Sigma_{cd}\rangle
&=-\eta_{[ab][cd]},
\qquad
\eta_{[ab][cd]}:=\eta_{ac}\eta_{bd}-\eta_{ad}\eta_{bc},
\label{eq:app-SigmaSigma}
\\
\langle \boldsymbol L_{ij}\boldsymbol L_{kl}\rangle
&= 2\left(\delta_{ik}\delta_{jl}-\delta_{il}\delta_{jk}\right),
\label{eq:app-LL}
\\
\langle \overline{\boldsymbol Q}^i_\alpha\,
\overline{\boldsymbol Q}^j_\beta\rangle
&=-2\,C_{\alpha \beta}\delta^{ij}.
\label{eq:app-QbarQbar}
\end{align}
Here and below the brackets mean the invariant matrix supertrace with the matrix ordering inherited from the corresponding term in Eq.~\eqref{LLlagrangian}.  Thus, for the frame-inserted terms, the ordered products that actually occur are, for example,
\begin{align}
\langle\se\wedge\FF\wedge*(\FF\wedge\se)\rangle
&=e^a\wedge F^A\wedge*(F^B\wedge e^b)\,
\operatorname{STr}
\!\left(\Gamma_a\boldsymbol T_A\boldsymbol T_B\Gamma_b\right),
\label{eq:app-ordered-4}
\\
\langle\se^2\wedge\FF\wedge*(\FF\wedge\se^2)\rangle
&=e^a\wedge e^b\wedge F^A\wedge
*(F^B\wedge e^c\wedge e^d)
\nonumber\\
&\quad\times
\operatorname{STr}
\!\left(\Gamma_{ab}\boldsymbol T_A\boldsymbol T_B
\Gamma_{cd}\right),
\label{eq:app-ordered-6}
\end{align}
where $\boldsymbol T_A$ runs over
$\{\boldsymbol\Sigma_{ab},\boldsymbol P_a,
\boldsymbol L_{ij},\boldsymbol Q_i,
\overline{\boldsymbol Q}{}^i\}$.  For the results of Table \ref{tab:SvN-ModGB}, this ordering should not be replaced, in these explicit matrix products, by an additional symmetrization of the four or
six matrices.

The elementary Clifford traces are
\begin{align}
\operatorname{Tr}(\mathbf 1)&=4,
&
\operatorname{Tr}(\gamma_a\gamma_b)&=4\eta_{ab},
\nonumber\\
\operatorname{Tr}(\gamma_{ab}\gamma_{cd})
&=-4\eta_{[ab][cd]},
&
\operatorname{Tr}(\hbox{odd number of }\gamma\hbox{'s})&=0.
\label{eq:app-basic-gamma-traces}
\end{align}
All higher even traces required in the calculation follow recursively from
\begin{equation}
\operatorname{Tr}(\gamma_{a_1}\cdots\gamma_{a_{2r}})
=
\sum_{j=2}^{2r}(-1)^j\eta_{a_1a_j}
\operatorname{Tr}
\!\left(
\gamma_{a_2}\cdots\widehat{\gamma}_{a_j}\cdots\gamma_{a_{2r}}
\right),
\label{eq:app-gamma-recursion-paper}
\end{equation}
where the hat denotes omission.  It is also useful to use
\begin{equation}
\gamma_a\gamma_b=\eta_{ab}+\gamma_{ab},
\qquad
[\boldsymbol\Sigma_{ab},\Gamma_c]
=\eta_{bc}\Gamma_a-\eta_{ac}\Gamma_b.
\label{eq:app-clifford-identities-paper}
\end{equation}

For the ordered frame--inserted products it is useful to introduce the normalized Dirac trace
\begin{equation}
\mathcal G_{a_1\cdots a_{2r}}
:=\frac14\operatorname{Tr}
\!\left(\gamma_{a_1}\cdots\gamma_{a_{2r}}\right),
\qquad
\mathcal G_{\varnothing}:=1.
\label{eq:app-G-kernel}
\end{equation}
Thus
\begin{equation}
\mathcal G_{ab}=\eta_{ab},
\qquad
\mathcal G_{abcd}
=\eta_{ab}\eta_{cd}-\eta_{ac}\eta_{bd}+\eta_{ad}\eta_{bc},
\label{eq:app-G2-G4}
\end{equation}
and Eq.~\eqref{eq:app-gamma-recursion-paper} becomes
\begin{equation}
\mathcal G_{a_1\cdots a_{2r}}
=
\sum_{j=2}^{2r}(-1)^j\eta_{a_1a_j}
\mathcal G_{a_2\cdots\widehat a_j\cdots a_{2r}}.
\label{eq:app-G-recursion}
\end{equation}
For products containing only the upper Clifford block the supertrace reduces
to the ordinary $4\times4$ Dirac trace.  Using
$\boldsymbol P_a=\Gamma_a/2$ and
$\boldsymbol\Sigma_{ab}=\tfrac14[\Gamma_a,\Gamma_b]$, the ordered
(unsymmetrized) supertraces that enter $\mathcal I_\beta$ are
\begin{align}
\left\langle
\Gamma_m\boldsymbol\Sigma_{ab}\boldsymbol\Sigma_{cd}\Gamma_n
\right\rangle
&=\frac14\Bigl(
\mathcal G_{mabcdn}-\mathcal G_{mbacdn}
-\mathcal G_{mabdcn}+\mathcal G_{mbadcn}
\Bigr)
\nonumber\\
&=-\eta_{mn}\eta_{[ab][cd]}
+\eta_{ma}\left(\eta_{bc}\eta_{dn}-\eta_{bd}\eta_{cn}\right)
\nonumber\\
&\quad
-\eta_{mb}\left(\eta_{ac}\eta_{dn}-\eta_{ad}\eta_{cn}\right)
+\eta_{mc}\left(\eta_{an}\eta_{bd}-\eta_{ad}\eta_{bn}\right)
\nonumber\\
&\quad
+\eta_{md}\left(\eta_{ac}\eta_{bn}-\eta_{an}\eta_{bc}\right),
\label{eq:app-ordered-SigmaSigma-4trace}
\\[2pt]
\left\langle
\Gamma_m\boldsymbol P_a\boldsymbol P_b\Gamma_n
\right\rangle
&=\mathcal G_{mabn}
=\eta_{ma}\eta_{bn}-\eta_{mb}\eta_{an}
+\eta_{mn}\eta_{ab}.
\label{eq:app-ordered-PP-4trace}
\end{align}
The corresponding ordered supertraces entering $\mathcal I_\gamma$ are
\begin{align}
\left\langle
\Gamma_{mn}\boldsymbol\Sigma_{ab}\boldsymbol\Sigma_{cd} \Gamma_{pq} \right\rangle
=&\frac{1}{16}\Bigl(\mathcal G_{mnabcdpq}-\mathcal G_{mnbacdpq}
-\mathcal G_{mnabdcpq}+\mathcal G_{mnbadcpq}\Bigr)\nonumber\\
&-\frac{1}{16}\Bigl(\mathcal G_{nmabcdpq}-\mathcal G_{nmbacdpq}
-\mathcal G_{nmabdcpq}+\mathcal G_{nmbadcpq}\Bigr)\nonumber\\
&-\frac{1}{16}\Bigl(\mathcal G_{mnabcdqp}-\mathcal G_{mnbacdqp}
-\mathcal G_{mnabdcqp}+\mathcal G_{mnbadcqp}\Bigr)\nonumber\\
&+\frac{1}{16}\Bigl(\mathcal G_{nmabcdqp}-\mathcal G_{nmbacdqp}
-\mathcal G_{nmabdcqp}+\mathcal G_{nmbadcqp}\Bigr)\label{eq:app-ordered-SigmaSigma-6trace}
\\[2pt]
\left\langle
\Gamma_{mn}\boldsymbol P_a\boldsymbol P_b\Gamma_{pq} \right\rangle
=&\frac14 \Bigl( \mathcal G_{mnabpq} - \mathcal G_{nmabpq} -\mathcal G_{mnabqp} + \mathcal G_{nmabqp}\Bigr).
\label{eq:app-ordered-PP-6trace}
\end{align}
No additional invariant tensors are hidden in these expressions.  In particular, the six-- and eight--index kernels in Eqs.~\eqref{eq:app-ordered-SigmaSigma-6trace} and \eqref{eq:app-ordered-PP-6trace} reduce completely to metric contractions:
\begin{align}
\mathcal G_{abcdef}
&=\eta_{ab}\mathcal G_{cdef}
-\eta_{ac}\mathcal G_{bdef}
+\eta_{ad}\mathcal G_{bcef}
-\eta_{ae}\mathcal G_{bcdf}
+\eta_{af}\mathcal G_{bcde},
\label{eq:app-G6-explicit}
\\
\mathcal G_{abcdefgh}
&=\eta_{ab}\mathcal G_{cdefgh}
-\eta_{ac}\mathcal G_{bdefgh}
+\eta_{ad}\mathcal G_{bcefgh}
-\eta_{ae}\mathcal G_{bcdfgh}
\nonumber\\
&\quad
+\eta_{af}\mathcal G_{bcdegh}
-\eta_{ag}\mathcal G_{bcdefh}
+\eta_{ah}\mathcal G_{bcdefg}.
\label{eq:app-G8-explicit}
\end{align}
Equations~\eqref{eq:app-ordered-SigmaSigma-4trace}--\eqref{eq:app-G8-explicit} are the unsymmetrized matrix ordering used in the Mathematica calculation: only the antisymmetrization already contained in $\boldsymbol\Sigma_{ab}$ and $\boldsymbol\Sigma_{cd}$ is performed; there is no further averaging over permutations of the matrices.

\subsection{Curvature decomposition in the notation of the main text}

The superconnection and curvature are, from Eqs.~\eqref{gaugeconnection} and \eqref{fieldstrength},
\begin{align}
\AAA
&=\frac12\omega^{ab}\boldsymbol\Sigma_{ab}
+e^a\boldsymbol P_a
+\frac12A^{ij}\boldsymbol L_{ij}
+\overline{\boldsymbol Q}{}^i_\alpha(\se\psi)^\alpha_i,
\label{eq:app-A-paper}
\\
\FF
&=\frac12F^{ab}\boldsymbol\Sigma_{ab}
+F^a\boldsymbol P_a
+\frac12F^{ij}\boldsymbol L_{ij}
+\overline{\boldsymbol Q}{}^i_\alpha X_i^\alpha.
\label{eq:app-F-paper}
\end{align}
The supercharge anticommutator is
\begin{equation}
\{\boldsymbol Q_i^\rho,
\overline{\boldsymbol Q}{}^j_\sigma\}
=
\left(
[\gamma^a]^\rho{}_\sigma\boldsymbol P_a
-\frac12[\gamma^{ab}]^\rho{}_\sigma
\boldsymbol\Sigma_{ab}
\right)\delta_i^j
+\delta^\rho_\sigma\boldsymbol L^j{}_i,
\label{eq:app-supercharge-anticommutator}
\end{equation}
so the bosonic curvature components contain the fermion bilinears displayed in
the main text,
\begin{align}
F^{ab}
&=R^{ab}+e^a\wedge e^b
-e^a\wedge e^b\,\bar\psi\psi
+*\!\left(e^a\wedge e^b\right)\bar\psi\gamma_*\psi,
\label{eq:app-Fab-full}
\\
F^a
&=T^a-*\!\left(e^a\wedge e^b\right)
\bar\psi\gamma_b\gamma_*\psi,
\label{eq:app-Fa-full}
\\
F^i{}_j
&=f^i{}_j-e^a\wedge e^b\,
\bar\psi^i\gamma_{ab}\psi_j,
\label{eq:app-Fij-full}
\\
f^i{}_j
&=dA^i{}_j+A^i{}_k\wedge A^k{}_j.
\label{eq:app-fij-paper}
\end{align}
The fermionic curvature is the one in Eq.~\eqref{Eq_X},
\begin{equation}
X_i
=-\se\wedge\mathrm D\psi_i
+\frac12\se^2\psi_i
+T^a\gamma_a\psi_i,
\qquad
\mathrm D=d+\frac12\omega^{ab}\Sigma_{ab}
+\frac12A^{mn}L_{mn}.
\label{eq:app-X-paper}
\end{equation}

For the coefficient map of Table~\ref{tab:SvN-ModGB} it is useful to separate three pieces.  The purely gravitational coefficients are obtained by setting
$\psi=0$ and $A^{ij}=0$, in which case
\begin{equation}
\hat{R}^{ab}:=\left.F^{ab}\right|_{\psi=0}=R^{ab}+e^a\wedge e^b,
\qquad
\left.F^a\right|_{\psi=0}=T^a.
\label{eq:app-pure-grav-curvatures}
\end{equation}
The pure Yang--Mills coefficient is obtained from $f^i{}_j$, while the pure fermionic kinetic and mass structures arise from the bilinear terms in $X_i$ and $\bar X^i$.  Contributions produced by inserting the fermionic corrections in Eqs.~\eqref{eq:app-Fab-full}--\eqref{eq:app-Fij-full} are mixed curvature--fermion, torsion--fermion, or four-fermion terms; these belong to $L_{\mathrm{NMC}}$ and are not displayed as independent rows of Table~\ref{tab:SvN-ModGB}.

We use the component conventions
\begin{equation}
R^{ab}=\frac12R^{ab}{}_{cd}\,e^c\wedge e^d,
\qquad
T^a=\frac12T^a{}_{bc}\,e^b\wedge e^c,
\qquad
f^i{}_j=\frac12f^i{}_{j\,ab}\,e^a\wedge e^b,
\label{eq:app-component-conventions}
\end{equation}
and
\begin{equation}
R^a{}_b:=R^{ac}{}_{bc},
\qquad
R:=R^a{}_a.
\end{equation}
For compactness in the intermediate steps, define
\begin{align}
\mathcal R_1&:=R_{abcd}R^{abcd},
&
\mathcal R_2&:=R_{abcd}R^{acbd},
&
\mathcal R_3&:=R_{abcd}R^{cdab},
\nonumber\\
\mathcal R_4&:=R_{ab}R^{ab},
&
\mathcal R_5&:=R_{ab}R^{ba},
\label{eq:app-R-basis}
\\
\mathcal T_1&:=T_{abc}T^{abc},
&
\mathcal T_2&:=T_{abc}T^{bca},
&
\mathcal T_3&:=T_b{}^{ba}T^c{}_{ca}.
\label{eq:app-T-basis}
\end{align}

\subsection{The six invariants}

To keep the calculation readable, denote the six terms in
Eq.~\eqref{LLlagrangian} by
\begin{align}
\mathcal I_\alpha
&:=\langle\FF\wedge*\FF\rangle,
\\
\mathcal I_\beta
&:=\langle\se\wedge\FF\wedge*(\FF\wedge\se)\rangle,
\\
\mathcal I_\gamma
&:=\langle\se^2\wedge\FF\wedge*(\FF\wedge\se^2)\rangle,
\\
\mathcal I_\delta
&:=\langle[\se,\FF]\wedge*[\FF,\se]\rangle,
\\
\mathcal I_\varepsilon
&:=\langle[\se^2,\FF]\wedge*[\FF,\se^2]\rangle,
\\
\mathcal I_\zeta
&:=\langle\se\wedge*\se\rangle.
\label{eq:app-six-invariants}
\end{align}
Thus
\begin{equation}
L_{\mathrm{LL}}
=\alpha\mathcal I_\alpha
+\beta\mathcal I_\beta
+\gamma\mathcal I_\gamma
+\delta\mathcal I_\delta
+\varepsilon\mathcal I_\varepsilon
+\zeta\mathcal I_\zeta.
\label{eq:app-LL-six}
\end{equation}

\subsection{Pure gravitational sector}

Set $\psi=0$ and $A^{ij}=0$.  Then
\begin{equation}
\left.\FF\right|_{\mathrm{grav}}
=\frac12\hat R^{ab}\boldsymbol\Sigma_{ab}
+T^a\boldsymbol P_a.
\label{eq:app-Fgrav-paper}
\end{equation}
It is useful first to give the complete reduction directly in differential-form
notation, without expanding the curvature and torsion into tensor components.
Let $\iota_a$ denote contraction with the frame vector dual to $e^a$.  The
Ricci one-form of the Lorentz component of the AdS curvature is
\begin{equation}
\mathfrak R^a:=\iota_b\hat R^{ba}.
\label{eq:app-AdS-Ricci-one-form}
\end{equation}
Indeed, using
\begin{equation}
\hat R^{ab}=\frac12\hat R^{ab}{}_{cd}\,
e^c\wedge e^d,
\end{equation}
one obtains
\begin{align}
\mathfrak R^a
&=\frac12\hat R^{ba}{}_{cd}\,
\iota_b\!\left(e^c\wedge e^d\right)
\nonumber\\
&=\frac12\hat R^{ba}{}_{cd}
\left(\delta_b^c e^d-\delta_b^d e^c\right)
\nonumber\\
&=\hat R^{ba}{}_{bc}\,e^c
=\hat R^{ab}{}_{cb}\,e^c
=\hat R^a{}_c\,e^c,
\qquad
\hat R^a{}_c:=\hat R^{ab}{}_{cb}.
\label{eq:app-AdS-Ricci-one-form-derivation}
\end{align}
Since $\hat R^{ab}=R^{ab}+e^a\wedge e^b$, this also gives
\begin{equation}
\hat R^a{}_c=R^a{}_c+3\delta^a_c,
\qquad
\mathfrak R^a=R^a{}_c e^c+3e^a.
\label{eq:app-AdS-Ricci-shift}
\end{equation}

The remaining differential forms needed in the reduction are
\begin{align}
\mathfrak B^a&:=\hat R^{ab}\wedge e_b,
&
\mathfrak C&:=e_a\wedge\mathfrak R^a,
\nonumber\\
\mathfrak A&:=e_a\wedge T^a,
&
\mathfrak t&:=\iota_aT^a.
\label{eq:app-derived-gravitational-forms}
\end{align}
Here $\mathfrak B^a$ is the curvature Bianchi three-form,
$\mathfrak C$ is the two-form associated with the antisymmetric part of the
Ricci tensor, while $\mathfrak A$ and $\mathfrak t$ are respectively the axial
three-form and trace one-form of the torsion.

Evaluating the ordered Clifford traces in
Eqs.~\eqref{eq:app-ordered-SigmaSigma-4trace}--\eqref{eq:app-ordered-PP-6trace}
and using the Hodge identity~\eqref{eq:app-hodge-basis-paper}, the six
gravitational invariants reduce to
\begin{align}
\left.\mathcal I_\alpha\right|_{\mathrm{grav}}
={}&-\frac12\hat R^{ab}\wedge*\hat R_{ab}
+T^a\wedge*T_a,
\label{eq:app-Ialpha-form-paper}
\\[2pt]
\left.\mathcal I_\beta\right|_{\mathrm{grav}}
={}&-\mathfrak B^a\wedge*\mathfrak B_a
-\mathfrak R^a\wedge*\mathfrak R_a
+T^a\wedge*T_a
+\mathfrak A\wedge*\mathfrak A
+\mathfrak t\wedge*\mathfrak t,
\label{eq:app-Ibeta-form-paper}
\\[2pt]
\left.\mathcal I_\gamma\right|_{\mathrm{grav}}
={}&\epsilon_{abcd}\hat R^{ab}\wedge\hat R^{cd}
-2\hat R^{ab}\wedge*\hat R_{ab}
+4\mathfrak B^a\wedge*\mathfrak B_a
+4\mathfrak R^a\wedge*\mathfrak R_a
\nonumber\\
&-4\mathfrak A\wedge*\mathfrak A
-4\mathfrak t\wedge*\mathfrak t,
\label{eq:app-Igamma-form-paper}
\\[2pt]
\left.\mathcal I_\delta\right|_{\mathrm{grav}}
={}&-4\mathfrak B^a\wedge*\mathfrak B_a
+4T^a\wedge*T_a
+4\mathfrak t\wedge*\mathfrak t,
\label{eq:app-Idelta-form-paper}
\\[2pt]
\left.\mathcal I_\varepsilon\right|_{\mathrm{grav}}
={}&16\mathfrak C\wedge*\mathfrak C
-16\mathfrak A\wedge*\mathfrak A,
\label{eq:app-Iepsilon-form-paper}
\\[2pt]
\mathcal I_\zeta
={}&16\vol.
\label{eq:app-Izeta-paper}
\end{align}
Thus no unevaluated traces of the form
$\langle\Gamma_m\boldsymbol T_A\boldsymbol T_B\Gamma_n\rangle$ or
$\langle\Gamma_{mn}\boldsymbol T_A\boldsymbol T_B\Gamma_{pq}\rangle$
remain in the gravitational reduction.

For comparison with the tensor basis used in Table~\ref{tab:SvN-ModGB}, the
relevant form identities are
\begin{align}
\hat R^{ab}\wedge*\hat R_{ab}
&=\left(\frac12\mathcal R_1+2R+12\right)\vol,
&
T^a\wedge*T_a&=\frac12\mathcal T_1\vol,
\nonumber\\
\mathfrak B^a\wedge*\mathfrak B_a
&=\left(\frac12\mathcal R_1-\mathcal R_2\right)\vol,
&
\mathfrak R^a\wedge*\mathfrak R_a
&=\left(\mathcal R_4+6R+36\right)\vol,
\nonumber\\
\mathfrak C\wedge*\mathfrak C
&=\left(\mathcal R_4-\mathcal R_5\right)\vol,
&
\mathfrak A\wedge*\mathfrak A
&=\left(\frac12\mathcal T_1+\mathcal T_2\right)\vol,
\nonumber\\
\mathfrak t\wedge*\mathfrak t
&=\mathcal T_3\vol,
\label{eq:app-form-to-component-identities}
\end{align}
together with
\begin{equation}
\epsilon_{abcd}\hat R^{ab}\wedge\hat R^{cd}
=\left(R^2+4R+24-4\mathcal R_5+\mathcal R_3\right)\vol.
\label{eq:app-AdS-Euler-component-identity}
\end{equation}
Substitution gives
\begin{align}
\left.\mathcal I_\alpha\right|_{\mathrm{grav}}
={}&-\vol\left(
R+6+\frac14\mathcal R_1-\frac12\mathcal T_1
\right),
\label{eq:app-Ialpha-paper}
\\
\left.\mathcal I_\beta\right|_{\mathrm{grav}}
={}&-\vol\left[
6R+36+\frac12\mathcal R_1-\mathcal R_2+\mathcal R_4
-\mathcal T_1-\mathcal T_2-\mathcal T_3
\right],
\label{eq:app-Ibeta-paper}
\\
\left.\mathcal I_\gamma\right|_{\mathrm{grav}}
={}&\vol\Bigl[
R^2+24R+144
+\mathcal R_1-4\mathcal R_2+\mathcal R_3
+4\mathcal R_4-4\mathcal R_5
\nonumber\\[-2pt]
&\hspace{21mm}-2\mathcal T_1-4\mathcal T_2-4\mathcal T_3
\Bigr],
\label{eq:app-Igamma-paper}
\\
\left.\mathcal I_\delta\right|_{\mathrm{grav}}
={}&-\vol\left[
2\mathcal R_1-4\mathcal R_2
-2\mathcal T_1-4\mathcal T_3
\right],
\label{eq:app-Idelta-paper}
\\
\left.\mathcal I_\varepsilon\right|_{\mathrm{grav}}
={}&\vol\left[
16\mathcal R_4-16\mathcal R_5
-8\mathcal T_1-16\mathcal T_2
\right].
\label{eq:app-Iepsilon-paper}
\end{align}
The $\delta$ and $\varepsilon$ expressions vanish in the torsionless sector
after imposing the standard algebraic Bianchi identities and therefore encode
torsion-dependent information in the general Riemann--Cartan geometry.

\subsection{Gauss--Bonnet identity}

For a metric-compatible Lorentz connection, the Euler four-form gives the Riemann--Cartan identity
\begin{equation}
\epsilon_{abcd}R^{ab}\wedge R^{cd} \;\longleftrightarrow\; R^2-4R_{ab}R^{ba}+R_{abcd}R^{cdab}.
\label{eq:app-Euler-paper}
\end{equation}
Hence, modulo the Euler number and boundary terms,
\begin{equation}
R^2\doteq4\mathcal R_5-\mathcal R_3.
\label{eq:app-GB-paper}
\end{equation}
Applying this identity to Eq.~\eqref{eq:app-Igamma-paper} gives the standard Gauss--Bonnet-reduced form
\begin{equation}
\left.\mathcal I_\gamma\right|_{\mathrm{grav}}^{\mathrm{GB}}
=\vol\left[
24R+144
+\mathcal R_1-4\mathcal R_2+4\mathcal R_4
-2\mathcal T_1-4\mathcal T_2-4\mathcal T_3
\right].
\label{eq:app-Igamma-GB-paper}
\end{equation}
This equation is the useful starting point whenever the quadratic gravitational sector is compared in a basis in which $R^2$ has genuinely been eliminated.

\subsection{Internal Yang--Mills sector}

For a purely internal curvature,
\begin{equation}
\left.\FF\right|_{so(n)}
=\frac12 f^{ij}\boldsymbol L_{ij},
\end{equation}
all frame-inserted invariants vanish because $\Gamma_a$ has support only on the upper Clifford block, whereas $\boldsymbol L_{ij}$ has support only on the lower $so(n)$ block.  Therefore
\begin{equation}
\left.\mathcal I_\alpha\right|_{so(n)}
=-\operatorname{Tr}(f\wedge*f)
=\frac12\vol\,f_{ij\,ab}f^{ij\,ab},
\label{eq:app-YM-paper}
\end{equation}
while
\begin{equation}
\left.\mathcal I_\beta\right|_{so(n)}
=\left.\mathcal I_\gamma\right|_{so(n)}
=\left.\mathcal I_\delta\right|_{so(n)}
=\left.\mathcal I_\varepsilon\right|_{so(n)}=0.
\end{equation}
Consequently the Yang--Mills row has coefficient $\alpha/2$.

\subsection{Fermionic sector}

The fermionic curvature carries the same internal index $i$ as the spinor in the superconnection,
\begin{equation}
X_i=-\se\wedge\mathrm D\psi_i
+\frac12\se^2\psi_i+T^a\gamma_a\psi_i.
\end{equation}
Let $\bar X^i$ denote its Majorana conjugate.  The terms in Table~\ref{tab:SvN-ModGB} which contain only $\psi_i$ and its covariant derivatives can be extracted by splitting
\begin{equation}
X_i=X_i^{(0)}+X_i^{(T)},
\qquad
X_i^{(0)}:=-\se\wedge\mathrm D\psi_i+\frac12\se^2\psi_i,
\qquad
X_i^{(T)}:=T^a\gamma_a\psi_i.
\label{eq:app-Xsplit-paper}
\end{equation}
Terms containing $X_i^{(T)}$ have an explicit torsion and are part of $L_{\mathrm{NMC}}$.

Define the two basic fermionic contractions
\begin{align}
\mathcal A_X
&:=\bar X^i\wedge*X_i,
\label{eq:app-AX-paper}
\\
\mathcal B_X
&:=\bar X^i\gamma_{ab}\wedge e^a\wedge*(X_i\wedge e^b).
\label{eq:app-BX-paper}
\end{align}
Direct multiplication of the odd blocks and use of
Eq.~\eqref{eq:app-QbarQbar} gives, for the pure-spinor structures retained in
Table~\ref{tab:SvN-ModGB},
\begin{align}
\left.\mathcal I_\alpha\right|_{\mathrm F}
&=2\mathcal A_X,
&
\left.\mathcal I_\beta\right|_{\mathrm F}
&=2\mathcal A_X+\mathcal B_X,
\nonumber\\
\left.\mathcal I_\gamma\right|_{\mathrm F}
&=-6\mathcal B_X,
&
\left.\mathcal I_\delta\right|_{\mathrm F}
&=4\mathcal A_X+2\mathcal B_X,
\nonumber\\
\left.\mathcal I_\varepsilon\right|_{\mathrm F}
&=-12\mathcal B_X.
\label{eq:app-fermion-map-paper}
\end{align}
Introduce exactly the four component structures listed in
Table~\ref{tab:SvN-ModGB},
\begin{align}
\mathcal K_D
&:=\bar\psi\gamma^\mu\mathrm D_\mu\psi
-\mathrm D_\mu\bar\psi\gamma^\mu\psi,
\label{eq:app-KD-paper}
\\
\mathcal M&:=\bar\psi\psi,
\\
\mathcal K_2
&:=\mathrm D^\mu\bar\psi\,\mathrm D_\mu\psi,
\\
\mathcal K_\gamma
&:=\mathrm D_\mu\bar\psi\,
\gamma^{\mu\nu}\mathrm D_\nu\psi.
\label{eq:app-Kgamma-paper}
\end{align}
The internal index is summed.  Expanding $X_i^{(0)}$ from Eq.~\eqref{eq:app-Xsplit-paper} in Eqs.~\eqref{eq:app-AX-paper} and \eqref{eq:app-BX-paper}, using $\se=e^a\gamma_a$ on the spinor block and applying Eq.~\eqref{eq:app-hodge-basis-paper}, gives
\begin{align}
\mathcal A_X
&=\vol\left(
3\mathcal K_D-3\mathcal M+3\mathcal K_2+\mathcal K_\gamma
\right)
+\hbox{terms containing }T^a,
\label{eq:app-AX-reduced-paper}
\\
\mathcal B_X
&=\vol\left(
12\mathcal K_D-12\mathcal M+6\mathcal K_2+6\mathcal K_\gamma
\right)
+\hbox{terms containing }T^a.
\label{eq:app-BX-reduced-paper}
\end{align}
Substituting these expressions into
Eq.~\eqref{eq:app-fermion-map-paper} gives
\begin{align}
\left.\mathcal I_\alpha\right|_{\mathrm F}
&=\vol\left(
6\mathcal K_D-6\mathcal M+6\mathcal K_2+2\mathcal K_\gamma
\right),
\nonumber\\
\left.\mathcal I_\beta\right|_{\mathrm F}
&=\vol\left(
18\mathcal K_D-18\mathcal M+12\mathcal K_2+8\mathcal K_\gamma
\right),
\nonumber\\
\left.\mathcal I_\gamma\right|_{\mathrm F}
&=-\vol\left(
72\mathcal K_D-72\mathcal M+36\mathcal K_2+36\mathcal K_\gamma
\right),
\nonumber\\
\left.\mathcal I_\delta\right|_{\mathrm F}
&=\vol\left(
36\mathcal K_D-36\mathcal M+24\mathcal K_2+16\mathcal K_\gamma
\right),
\nonumber\\
\left.\mathcal I_\varepsilon\right|_{\mathrm F}
&=-\vol\left(
144\mathcal K_D-144\mathcal M+72\mathcal K_2+72\mathcal K_\gamma
\right).
\label{eq:app-fermion-final-paper}
\end{align}
The coefficients of $\mathcal K_D$, $\mathcal M$, $\mathcal K_2$, and
$\mathcal K_\gamma$ therefore reproduce the four fermionic rows of
Table~\ref{tab:SvN-ModGB}.

\subsection{Collection of the coefficient map}

Combining Eqs.~\eqref{eq:app-Ialpha-paper},
\eqref{eq:app-Ibeta-paper}, \eqref{eq:app-Igamma-paper},
\eqref{eq:app-Idelta-paper}, \eqref{eq:app-Iepsilon-paper},
\eqref{eq:app-YM-paper}, and \eqref{eq:app-fermion-final-paper}, the direct
component expansion before eliminating $R^2$ is
\begin{align}
\frac{L_{\mathrm{LL}}}{\vol}
={}&(-\alpha-6\beta+24\gamma)R
+(-6\alpha-36\beta+144\gamma+16\zeta)
+\gamma R^2
\nonumber\\
&+\left(-\frac\alpha4 -\frac\beta2+\gamma-2\delta\right)\mathcal R_1
+(\beta-4\gamma+4\delta)\mathcal R_2
+\gamma\mathcal R_3
\nonumber\\
&+(-\beta+4\gamma+16\varepsilon)\mathcal R_4
+(-4\gamma-16\varepsilon)\mathcal R_5
\nonumber\\
&+\left(\frac\alpha2+\beta-2\gamma+2\delta-8\varepsilon\right)\mathcal T_1
+(\beta-4\gamma-16\varepsilon)\mathcal T_2
\nonumber\\
&+(\beta-4\gamma+4\delta)\mathcal T_3
+\frac\alpha2 f_{ij\,ab}f^{ij\,ab}
\nonumber\\
&+(6\alpha+18\beta-72\gamma+36\delta-144\varepsilon)\mathcal K_D
\nonumber\\
&-(6\alpha+18\beta-72\gamma+36\delta-144\varepsilon)\mathcal M
\nonumber\\
&+(6\alpha+12\beta-36\gamma+24\delta-72\varepsilon)\mathcal K_2
\nonumber\\
&+(2\alpha+8\beta-36\gamma+16\delta-72\varepsilon)\mathcal K_\gamma
+\frac{L_{\mathrm{NMC}}}{\vol}.
\label{eq:app-master-paper}
\end{align}

Equation~\eqref{eq:app-master-paper} is the direct result of the matrix-block calculation before eliminating $R^2$. Using Eq.~\eqref{eq:app-GB-paper},
\[
\gamma\left(
R^2+\mathcal R_3+4\mathcal R_4-4\mathcal R_5
\right)
\doteq
4\gamma\mathcal R_4 .
\]
Thus the $R_{abcd}R^{cdab}$ term and the $\gamma$-dependent contribution to $R_{ab}R^{ba}$ cancel, while the $\gamma$-dependent contribution to the coefficient of $R_{ab}R^{ab}$ is $4\gamma$.

\subsection{Reduction of the antisymmetric two-derivative fermion term}
\label{app:Dgamma2D}

For completeness, we show explicitly that the antisymmetric two-derivative structure
\begin{equation}
\mathcal K_\gamma
:=
\mathrm D_\mu\bar\psi^i\,
\gamma^{\mu\nu}\mathrm D_\nu\psi_i
\end{equation}
does not constitute an independent second-order fermionic kinetic term, as mentioned in \eqref{Eq_Term_DpsiGamma_abDpsi}. Here $\mathrm D_\mu$ denotes the Lorentz- and $SO(n)$-covariant derivative introduced in the main text. Let $\widehat{\mathrm D}_\mu$ denote its extension to spacetime tensor indices. In particular,
\begin{equation}
\widehat{\mathrm D}_\mu
\left(\mathrm D_\nu\psi_i\right)
=
\mathrm D_\mu\mathrm D_\nu\psi_i
-\Gamma^\rho{}_{\mu\nu}\mathrm D_\rho\psi_i,
\end{equation}
and metric compatibility implies
\begin{equation}
\widehat{\mathrm D}_\rho\gamma^{\mu\nu}=0.
\end{equation}

Defining
\begin{equation}
V^\mu
:=
\bar\psi^i\gamma^{\mu\nu}\mathrm D_\nu\psi_i,
\end{equation}
we have
\begin{align}
\mathcal K_\gamma
&=
\widehat{\nabla}_\mu V^\mu
-
\bar\psi^i\gamma^{\mu\nu}
\widehat{\mathrm D}_\mu\mathrm D_\nu\psi_i
\nonumber\\
&=
\widehat{\nabla}_\mu V^\mu
-\frac12
\bar\psi^i\gamma^{\mu\nu}
[\widehat{\mathrm D}_\mu,\widehat{\mathrm D}_\nu]\psi_i,
\label{eq:app-Kgamma-ibp}
\end{align}
where the second equality follows from the antisymmetry of
$\gamma^{\mu\nu}$.

For a spinor in Riemann--Cartan geometry the commutator is
\begin{equation}
[\widehat{\mathrm D}_\mu,\widehat{\mathrm D}_\nu]\psi_i
=
\frac14R_{\mu\nu}{}^{ab}\gamma_{ab}\psi_i
+
(f_{\mu\nu})_i{}^j\psi_j
-
T^\rho{}_{\mu\nu}\mathrm D_\rho\psi_i.
\label{eq:app-spinor-commutator-torsion}
\end{equation}
Consequently,
\begin{align}
\mathcal K_\gamma
={}&
\widehat{\nabla}_\mu V^\mu
-\frac18R_{\mu\nu}{}^{ab}
\bar\psi^i\gamma^{\mu\nu}\gamma_{ab}\psi_i
-\frac12
\bar\psi^i\gamma^{\mu\nu}(f_{\mu\nu})_i{}^j\psi_j
\nonumber\\
&+
\frac12T^\rho{}_{\mu\nu}
\bar\psi^i\gamma^{\mu\nu}\mathrm D_\rho\psi_i.
\label{eq:app-Kgamma-before-boundary}
\end{align}

Introducing the torsion trace
\begin{equation}
t_\mu:=T^\rho{}_{\mu\rho},
\end{equation}
the divergence of a vector in a metric-compatible Riemann--Cartan geometry satisfies
\begin{equation}
\widehat{\nabla}_\mu V^\mu
=
\frac{1}{\sqrt{|g|}}
\partial_\mu\left(\sqrt{|g|}V^\mu\right)
-t_\mu V^\mu.
\end{equation}
Dropping the total derivative, Eq.~\eqref{eq:app-Kgamma-before-boundary} therefore becomes
\begin{align}
\mathcal K_\gamma
\doteq{}&
-\frac18R_{\mu\nu}{}^{ab}
\bar\psi^i\gamma^{\mu\nu}\gamma_{ab}\psi_i
-\frac12
\bar\psi^i\gamma^{\mu\nu}(f_{\mu\nu})_i{}^j\psi_j
\nonumber\\
&+
\frac12T^\rho{}_{\mu\nu}
\bar\psi^i\gamma^{\mu\nu}\mathrm D_\rho\psi_i
-
t_\mu\bar\psi^i\gamma^{\mu\nu}\mathrm D_\nu\psi_i,
\label{eq:app-Kgamma-final}
\end{align}
where $\doteq$ denotes equality up to a boundary term. Thus $\mathcal K_\gamma$ can be traded for nonminimal couplings to the Lorentz curvature and internal field strength, together with torsion-dependent first-order fermionic couplings, and does not provide an independent second-order kinetic operator.

\section{Contortion expansion and linearized interpretation}\label{app:linearizedtorsionmodes}

Decompose the spin connection into Levi--Civita part plus contortion,
\begin{equation}
\omega^{ab}=\overset{\circ}{\omega}{}^{ab}(e)+\kappa^{ab},
\qquad
T^a := \mathrm{D}e^a = \kappa^a{}_b\wedge e^b,
\label{eq:omega_split}
\end{equation}
where $\overset{\circ}{\omega}{}^{ab}(e)$ is determined by $e^a$ and has vanishing torsion.
The Lorentz curvature expands as
\begin{equation}
R^{ab}(\omega)
=
R^{ab}(\overset{\circ}{\omega})
+\overset{\circ}{\mathrm{D}}\kappa^{ab}
+\kappa^a{}_c\wedge\kappa^{cb},
\label{eq:F_expand}
\end{equation}
with $\overset{\circ}{\mathrm{D}}$ the covariant derivative built from $\overset{\circ}{\omega}$.
Linearizing about Minkowski space,
\begin{equation}
e^a{}_\mu=\delta^a_\mu + \frac12\,h^a{}_\mu,\qquad
\kappa^{ab}{}_\mu=\mathcal{O}(\epsilon),
\end{equation}
one finds to first order in perturbations:
\begin{equation}
R^{ab} \;\approx\; R^{ab}(\overset{\circ}{\omega})
+\overset{\circ}{\mathrm{D}}\kappa^{ab},
\qquad
\kappa\wedge \kappa=\mathcal{O}(\epsilon^2).
\end{equation}

\subsection{Spin--parity content}\label{app:AS_to_spinparity}
In Poincar\'e gauge theory, the propagating modes are classified by decomposing torsion and curvature into Lorentz-irreducible pieces and then using spin projection operators. For our purposes, it is enough to recall the standard decomposition of torsion in 4D into
\begin{equation}
T_{\lambda\mu\nu} = \frac{1}{3}\left(v_\mu g_{\lambda\nu}-v_\nu g_{\lambda\mu}\right)
+ \frac{1}{6}\,\varepsilon_{\lambda\mu\nu\rho}\,a^\rho 
+ q_{\lambda\mu\nu},
\label{eq:torsion_decomp}
\end{equation}
where $v_\mu:=T^\lambda{}_{\mu\lambda}$ is the trace vector (parity even),
$a^\mu:=\varepsilon^{\mu\alpha\beta\gamma}T_{\alpha\beta\gamma}$ is the axial vector (parity odd),
and $q_{\lambda\mu\nu}$ is the remaining tensor piece (traceless and $\varepsilon$-traceless).

The spin--parity identification proceeds by inserting \eqref{eq:torsion_decomp} into the quadratic action and decomposing the resulting fields into their irreducible spin components. The propagating
sectors are then identified from the corresponding kinetic operators:
\begin{itemize}
 \item The $\mathbf{2}^+$ sector contains the usual massless graviton associated with the tetrad (or metric). Its standard kinetic term is already present in the Einstein--Hilbert contribution $R(\overset{\circ}{\omega})$.
 \item The torsion trace vector $v_\mu$ decomposes into the spin--parity sectors: $\mathbf{0}^+\oplus\mathbf{1}^-$. More precisely, its longitudinal component carries the $\mathbf{0}^+$ sector, whereas its transverse component carries the $\mathbf{1}^-$ sector. Since derivatives of $v_\mu$ originate from the $\overset{\circ}{\mathrm D}\kappa$ contribution to the curvature, kinetic terms for these torsional modes arise from the curvature-quadratic invariants.

 \item The axial pseudovector $a_\mu$ similarly decomposes as $\mathbf{0}^-\oplus\mathbf{1}^+$. Its longitudinal component carries the pseudoscalar $\mathbf{0}^-$ mode, while its transverse component carries the $\mathbf{1}^+$ mode. Their kinetic terms likewise originate from the $\overset{\circ}{\mathrm D}\kappa$ part of the curvature entering the quadratic curvature invariants.

 \item The tensor component $q_{\lambda\mu\nu}$ contains the remaining irreducible spin sectors of the torsion. Whether any of these sectors propagates depends on the particular combination of quadratic invariants and on the resulting degeneracies of the kinetic operator.
\end{itemize}
Which of these sectors actually propagates, and whether it is healthy, is determined by the pole structure and the signs of the corresponding residues in the quadratic action.

Finally, linear perturbations of Riemann--Cartan geometry can be analyzed without abandoning the first--order differential--forms formalism \eqref{eq:torsion_decomp}, and without resorting to an \emph{a priori} splitting between torsionless and torsional components of the connection \eqref{eq:omega_split}; see refs.~\cite{Izaurieta:2019dix,Barrientos:2019msu,Barrientos:2019awg,Elizalde:2022vvc,Barriga:2024hpe}. These methods provide a fully geometric framework for the perturbative analysis of Riemann--Cartan theories. In future work, they will be employed to determine the spin--parity content of the propagating modes and to perform a systematic identification of ghost degrees of freedom on more general backgrounds.

\end{appendices}

\bibliographystyle{ieeetr}
\bibliography{clifford-algebra-gravity-gut.bib}

\end{document}